\documentclass[preprint,letterpaper,12pt]{JHEP3}
\usepackage{axodraw}
\usepackage{graphics,epsfig}
\newcommand{\GeV}{\,{\rm GeV}}
\newcommand{\MeV}{\,{\rm MeV}}

\newcommand\lsim{\mathrel{\rlap{\lower4pt\hbox{\hskip1pt$\sim$}}
    \raise1pt\hbox{$<$}}}
\newcommand\gsim{\mathrel{\rlap{\lower4pt\hbox{\hskip1pt$\sim$}}
    \raise1pt\hbox{$>$}}}
\def\bea{\begin{eqnarray}}
\def\eea{\end{eqnarray}}
\def\ba{\begin{array}}
\def\ea{\end{array}}
\def\bec{\begin{center}}
\def\ec{\end{center}}
\def\nn{\nonumber}

\def\64{\rm SO(6) \times SO(4)}

\def\f{\frac}
\def\a{\alpha}
\def\b{\beta}

\def\e{\epsilon}
\def\ve{\varepsilon}
\def\g{\gamma}

\def\l{\lambda}
\def\t{\theta}
\def\z{\zeta}

\def\[{\left[}
\def\]{\right]}
\def\({\left(}
\def\){\right)}

\preprint{OHSTPY-HEP-T-04-013 \\  SNUTP 04-018}
\title{\huge Quark and Lepton Masses in 5D SO(10)}
\author{Hyung Do Kim$^{a,b}$, Stuart Raby$^a$ and Leslie Schradin$^a$\\
$^a$Department of Physics, The Ohio State University,\\
174 W. 18th Ave., Columbus, Ohio 43210, USA\\
$^b$School of Physics, Seoul National University,\\
Seoul, 151-747, Korea\\
\\ E-mail: \email{hdkim@phya.snu.ac.kr \\
raby,schradin@mps.ohio-state.edu}}

\abstract{ We construct a five dimensional supersymmetric SO(10)$\times$D$_3$ grand unified model with an
$S^1/\left( Z_2 \times Z^\prime_2 \right)$ orbifold as the extra dimension. The orbifold breaks half of the
supersymmetry and breaks the SO(10) gauge symmetry down to ${\rm SU(4)}_C \times {\rm SU(2)}_L \times {\rm
SU(2)}_R$. The Higgs mechanism is used to break the remaining gauge symmetry the rest of the way to the Standard
Model. We place matter fields variously in the bulk and on the orbifold fixed points and the resulting massless
fields are mixtures between these brane and bulk fields. A chiral adjoint field in the bulk gets a U(1)$_X$
vacuum expectation value, resulting in an $X$-dependent localization of the bulk matter fields and the Standard
Model Higgs field. This Higgs field localization allows us to simultaneously explain the hierarchies $m_u < m_d$
and $m_t \gg m_b$. The model uses 11 parameters to fit the 13 independent low energy observables of the quark and
charged lepton Yukawa matrices. The model predicts the values of two quark mass combinations, $\f{m_u}{m_c}$ and
$m_d m_s m_b$, each of which are predicted to be approximately $1 \sigma$ above their experimental values. The
remaining observables are successfully fit at the 5\% level.  We note that this 5D theory, as formulated, has
problems retaining the Pati-Salam Yukawa symmetry relations.  A simple 6D fix, which preserves the 5D results,
solves this problem.}


\begin{document}

\section{Introduction}

The Standard Model (SM) of particle physics has had spectacular success
in describing the strong, weak, and electromagnetic forces of
nature in terms of gauge theory. It does, however, leave several
issues unexplained. Among these issues: First, in the Standard
Model the three gauge coupling values are arbitrary and unrelated,
but appear to almost unify at a high scale. The assumption of
supersymmetry improves this unification. Second, the fermion
charge assignments under the gauge symmetries are arbitrary.
Third, recent experiments have shown that neutrinos have masses,
but they are massless in the Standard Model. Fourth, in a universe
initially balanced between matter and anti-matter, interactions
which violate baryon number are required to explain the apparent
baryon asymmetry observed in the universe. Baryon number is conserved by the
renormalizable terms of the Standard Model.
Fifth, the weak and Planck scales are 16
orders of magnitude apart, causing a naturalness problem in the
mass of the Higgs boson. While this is not an exhaustive list of
problems with the Standard Model, all of these issues are
addressed by four dimensional supersymmetric grand unified
theories [4D SUSY  GUTS]
\cite{Dimopoulos:1981yj,Dimopoulos:1981zb,Ibanez:yh,Sakai:1981gr,Einhorn:1981sx,Marciano:1981un}.

In a SUSY GUT, the three gauge couplings meet near the GUT scale $M_G \simeq 3 \times 10^{16} \GeV$,
and above this scale an enlarged symmetry group with a single gauge coupling governs physics.
Within this paper, we will take SO(10) as our gauge group. Remarkably, the charges of the SM fermions
are exactly those which come from a 16 of SO(10), including the arbitrary (in the SM) U(1)$_Y$
hypercharge quantum numbers. Moreover, a 16 dimensional representation requires the addition of a sterile neutrino
which can lead to tiny neutrino masses through a See-Saw mechanism.
$B-L$ is a gauge symmetry within SO(10) which is broken at a high scale
in the unified theory, thus making Baryogenesis possible.
Finally, the presence of supersymmetry near the weak
scale gives a natural explanation for a Higgs mass at that scale.

As experiments become more accurate, we are able to test whether GUT theories work
in their minimal forms. There are several indications that the most
simple GUT theories cannot work.
\begin{itemize}
\item
With the definition of the GUT scale $M_G$: $\a_1(M_G) \equiv \a_2(M_G) \equiv \a_{\rm GUT}$, the assumption of exact unification
($\a_3(M_G) = \a_{\rm GUT}$) leads to a prediction for $\a_3(M_Z)$ which is larger than the measurements.
This indicates that a negative threshold correction to $\a_3$ must
be present at the GUT scale: $\varepsilon_3 \equiv \f{\a_3(M_G) - \a_{\rm GUT} }{\a_{\rm GUT}} \simeq -0.04$ \cite{Raby:er}.

\item
Since the fermions are combined into fewer multiplets of the unifying group,
GUT theories in general give relationships between the fermion masses at the GUT scale.
In SO(10), third family unification can be accomodated $(m_\tau(M_G) = m_b(M_G) = m_t(M_G))$, but the
first and second family masses do not unify at $M_G$.
The Georgi-Jarlskog relation: $m_s/m_\mu = \f{1}{3} m_b/m_\tau$ \cite{Georgi:1979df},
which had been known to work very well,
holds less well as measurements of the strange quark mass decrease.

\item
GUT models place the weak Higgs doublet(s) into larger representations which
include new color triplet Higgs fields. It is difficult to make the color triplets
heavy enough (near GUT scale mass) to avoid rapid proton decay while keeping
the weak doublets light (weak scale mass). The doublet-triplet (DT) splitting
problem is related to the method by which the GUT symmetry is broken to the Standard
Model, and the models which give proper DT splitting can be quite complicated.
\end{itemize}

Orbifold GUTS have been considered extensively for the last five years \cite{Kawamura:2000ev, Kawamura:2000ir,
Altarelli:2001qj, Hall:2001pg, Hall:2001xb, Hall:2002ci, Kobakhidze:2001yk,
Hebecker:2001wq, Hebecker:2001jb, Dermisek:2001hp, Asaka:2001eh}
and provide answers to some of the above issues.
If the compactification radius is slightly smaller than the cutoff scale,
the heavy Kaluza-Klein (KK) modes can give compactification scale threshold corrections necessary for exact gauge unification
at the cutoff scale \cite{Kim:2002im}. If the Standard Model Higgs is a bulk field,
and if the triplet Higgs has twisted boudary conditions under the orbifold
it will acquire a mass at the compactification scale,
thus solving the DT splitting problem naturally. The same mechanism breaks the GUT gauge symmetry
and gives mass to gauge bosons outside of the Standard Model.

5D SU(5) models have a SM brane on which all of the interesting GUT mass relations are absent. The first and
second family matter fields usually feel this brane and so there are no relations between quarks and leptons in
these two families. This is unattractive since the models lose predictivity. For this reason, we choose to
concentrate on SO(10) models which give more possible avenues of symmetry breaking to the Standard Model and
allow for GUT mass relations.

5D SO(10) SUSY GUTS were first considered in Dermisek and Mafi \cite{Dermisek:2001hp}, and a setup which gives
gauge coupling unification was constructed by Kim and Raby \cite{Kim:2002im}. The extra dimension is an $S^1 /
(Z_2 \times Z_2^\prime)$ orbifold, a line segment with endpoints which are fixedpoints of the orbifold $Z_2$
symmetries. 5D N=1 SUSY is broken to 4D N=1 by the first $Z_2$, while the second $Z_2$ breaks SO(10) to SO(6)
$\times$ SO(4) ($=$ SU(4) $\times$ SU(2) $\times$ SU(2) $\equiv$ Pati-Salam gauge group). Further breaking to the
Standard Model is provided by the Higgs mechanism on one of the fixed points. One fixed point, the SO(10) brane,
is invariant under the first $Z_2$ and has SO(10) gauge symmetry. The other fixed point, the Pati-Salam (PS)
brane, is invariant under the second $Z_2$ and only has Pati-Salam gauge symmetry. The bulk has the full SO(10)
gauge symmetry. Gauge coupling unification works if the compactification scale is approximately $10^{14} \GeV$,
the cutoff scale is approximately $10^{17} \GeV$, and if the Higgs multiplet lives in the bulk.  A virtue of the
model is that SO(10) or Pati-Salam exists after the orbifold breaking and we retain some of the Yukawa GUT
relations.

We note that there is a problem with the five dimensional formulation of our theory. In order for the gauge
couplings to unify at the cutoff scale, the PS brane fields breaking Pati-Salam down to the Standard Model must
have cutoff-scale VEVs. Therefore, in general one would expect all aspects of PS symmetry to be broken at this
high scale, and that the lower-energy theory should not be expected to exhibit PS symmetry. Nevertheless, we wish
to retain some of the PS symmetry below the cutoff scale in order to preserve the PS Yukawa coupling relations.
This is possible if there is an additional extra dimension in which the PS brane is separated from the
PS-breaking VEV fields. This 6th direction will solve this problem if its length scale is only slightly larger
than the cutoff scale. The effective 5D theory below the energy scale of this 6th direction will not be affected
in any other way. For this reason we choose to perform our analysis in the 5D theory, and will assume that PS
Yukawa relations are still valid below the cutoff-scale. We discuss the 6D fix to the PS-breaking problem in
Section \ref{sec:6D}.

Attempts to explain the mass hierarchy between the different families of fermions have often
centered on flavor symmetries.
There are models based on abelian horizontal U(1) symmetries where the
smallness of certain couplings is explained by the suppression given by high powers of
U(1) breaking fields. The explanations can at most be qualitative as there are order
one coefficients which are undetermined from the U(1) symmetry. We therefore choose to concentrate on
non-abelian flavor symmetry.
With three families, the largest possible symmetry would be SU(3).
However, the order one top Yukawa coupling would badly break the SU(3) symmetry, and
so instead we choose to concentrate on an SU(2) symmetry between the first two families.
This symmetry can explain the absence of flavor changing neutral currents in supersymmetric theories
and can relate unknown order one coefficients between different families.
In string theory, global symmetries are thought to be generally broken by quantum gravity
effects \cite{Giddings:1988cx, Coleman:1988tj, Gilbert:1989nq}, and so
our flavor symmetry should be a gauge symmetry. However, the breaking of a continous
gauge symmetry like SU(2) gives unwanted flavor changing neutral currents from the
D-term contributions \cite{Kawamura:1994ys, Babu:1999js}.
For this reason, we assume a flavor symmetry
relating the first and second families based on D$_3$, a discrete subgroup of SU(2).
As a discrete gauge symmetry, D$_3$ has all the virtues of SU(2)
and can avoid the problem related to continuous gauge symmetry \cite{Krauss:1988zc}.
More information and references on family symmetries in grand unified theories can be found
in the reviews \cite{Albright:2002np, Chen:2003zv, Altarelli:2004za}.

In SUSY SO(10) models with flavor symmetries, the most difficult thing to explain is why $m_u < m_d$ while $m_t
\gg m_b$. In SO(10) models, we usually have $m_t/m_b = Y_t/Y_b \, \tan \b \sim \tan \b$, and the heaviness of the
top quark is explained by a large value for $\tan \b \sim 50$. In these models, it is natural that $Y_u/Y_d \sim
Y_t/Y_b$, which leads to $m_u/m_d \sim 50$. To fit the data, an unusually small Yukawa coupling is needed for the
up quark compared to the down quark. This is difficult to implement in SO(10) models. Extra-dimensional theories
provide a nice tool to suppress or enhance fermion masses, as the size of a mass can be influenced by the
localization of the matter and Higgs fields within the extra dimension. Others, notably Arkani-Hamed and Schmaltz
in \cite{Arkani-Hamed:1999dc}, have used localization in an extra dimension to explain the fermion mass
hierarchy. Our model, on the other hand, retains some of the attractive Yukawa relations present in 4D GUT models
which are not present in \cite{Arkani-Hamed:1999dc}.

We explain the ratios $m_u/m_d < 1$ and $m_t/m_b \gg 1$
simultaneously by the quasi-localization of the Higgs field in the
bulk by a kink mass.\footnote{The extreme limit is discussed in
\cite{Kim:2001at} where $H_u$ is on the brane and $H_d$ is in the
bulk such that $m_t \gg m_b$ for order one $\tan \beta$.} We give
a vacuum expectation value (VEV) in the U$(1)_X$ direction to a
scalar adjoint field $\Sigma$ contained in the 5D vector
multiplet. $\Sigma$ is odd under both $Z_2$ parities, and we
choose the VEV so that its contribution to the D-term VEV is only
at the two branes. Supersymmetry can be preserved if there are
fields on each brane which get VEVs in the right-handed neutrino
direction to counter the contribution of $\left< \Sigma \right>$
to the D-term VEV. The VEV of $\Sigma$ acts to give a
U$(1)_X$-dependent mass to the hypermultiplet fields in the bulk,
in particular the Higgs field. Because $H_u$ and $H_d$ carry
opposite U$(1)_X$ charges, they are localized towards different
branes. Thus if the 1st and 3rd families get their Yukawa
couplings on opposite branes, we can naturally have $m_u/m_d < 1$
and $m_t/m_b > 1$.

The easiest way to ensure that the 1st and 3rd families get their Yukawa couplings on opposite branes is to
restrict them to these branes. By proton decay constraints, the 1st family cannot reside on the SO(10) brane.
Thus we choose the 1st (and by family symmetry, the 2nd) family to be on the PS brane, and the 3rd family to be
on the SO(10) brane. We then choose the sign of the kink mass VEV to be such that $H_u$ is localized towards the
SO(10) brane and $H_d$ is localized towards the PS brane.\footnote{These arguments do not rule out the case of
all 3 families in the bulk, or some fields on branes with others in the bulk. However, if the 3rd family is a
brane field, then the 2nd family must also be a brane field or the ratio $\f{m_\mu}{m_\tau}$ is made too small by
volume suppression to fit the data. It is possible to decrease the suppression from the volume factor by either
placing only half of the 2nd family in the bulk and/or by altering the GUT threshold correction $\ve_3$. We made
some attempts with these approaches but were unable to produce viable theories.} Communication between the first
two families and the third is provided by mixing with bulk fields. After requiring that the 1st and 3rd families
be brane fields, the location of the Higgs and all matter fields is entirely determined from gauge coupling
unification, proton decay constraints, D$_3$ family symmetry between the first two families, and the action of
the kink mass to explain 1st and 3rd family mass ratios.

To summarize: In this paper, we construct a 5D SO(10) supersymmetric orbifold GUT model in which the 1st and 2nd
families reside on the PS brane and transform as a doublet under the D$_3$ family symmetry, the 3rd family is
located on the SO(10) brane, and a kink mass localizes the MSSM Higgs doublets to opposite sides of the bulk.
The model explains the 13 independent quark and charged lepton
masses and mixing angles in terms of 11 parameters.
The two predictions are
$\f{m_u}{m_c} (M_Z) = 0.0037 \pm 0.0006$ and $m_d m_s m_b (M_Z) = (10.7 \pm 5.0) \times 10^5 \MeV^3$,
both of which are approximately $1 \sigma$ above the experimental values.

We lay out the remainder of this paper as follows. The data is summarized in Section \ref{sec:Data}. The basic
setup is explained in Section \ref{sec:Setup}. This includes background material and our choice of superpotential
and Yukawa matrices. Section \ref{sec:Analysis} contains the analysis of our data. This part is broken into two
subsections: an analytic section (Section \ref{sec:Analytic}) giving predictions and approximate relationships
between the model parameters and the data, and a numerical section (Section \ref{sec:Numerical}) with more
precise results. The possibilities and advantages of placing our model in a 6D framework are given in Section
\ref{sec:6D}. Finally, we present a summary and discussion of our model in Section \ref{sec:Summary}. We also
provide necessary information on D$_3$ family symmetry in Appendix \ref{sec:AppD_3} and a determination of
massless state wavefunctions in Appendix \ref{sec:AppWave}.  Finally, in Appendix \ref{sec:App4DModel} we
consider a 4D SO(10) SUSY GUT model \cite{Dermisek:1999vy} lifted up to 5D. In this model the smallness of the up
quark mass is explained by an approximate left-right (LR) symmetry rather than by a kink VEV.

\section{Data \label{sec:Data}}

In this section we tabulate the low energy data used in our analysis.
First the data associated with the Cabbibo-Kobayashi-Maskawa (CKM) matrix. These CKM elements we take from \cite{Battaglia:2003in}.
\bea
\left| V_{us} \right| & = & 0.2240 \pm 0.0036 \\
\left| V_{cb} \right| & = & (41.5 \pm 0.8) \times 10^{-3} \nn \\
\left| \f{V_{ub}}{V_{cb}} \right| & = & 0.086 \pm 0.008 \nn
\eea
$|V_{td}|$ and $\sin 2 \beta$ we take from \cite{Hocker:2001xe}.
\bea
\left| V_{td} \right| & = & 0.0082 \pm 0.0008 \\
\sin 2 \b & = & 0.739 \pm 0.048 \nn
\eea
We use \cite{Hagiwara:fs} for $J$ and $\varepsilon_K$.
\bea
J & = & (3.0 \pm 0.3) \times 10^{-5} \\
\varepsilon_K & = & (2.282 \pm 0.017) \times 10^{-3} \nn
\eea

Next the low energy quark mass observables.
We use
\bea
Q \equiv \f{\f{m_s}{m_d}}{\sqrt{1-\left( \f{m_u}{m_d} \right)^2}}
\eea
and $m_s/m_d$ from \cite{Leutwyler:1996qg} with doubled errors on $m_s/m_d$,
the unquenched lattice QCD result with $n_f =2$ for $m_s$ \cite{Gupta:2001cu} with doubled errors,
and we take the quenched lattice QCD result for $m_c$ with 10\% error from \cite{Battaglia:2003in} and double the error.
We use the bottom quark mass from \cite{Bauer:2002sh} and the top quark pole mass from the
CDF and D$\emptyset$ Collaboration \cite{Azzi:2004rc}.
\bea
Q & = & 22.7 \pm 0.8 \\
\f{m_s}{m_d} & = & 18.9 \pm 0.8 \times 2 \nn \\
m_s (2\GeV) & = & 89 \pm 11 \times 2 \MeV \nn \\
m_c (m_c) & = & 1.30 \pm 0.15 \times 2 \GeV \nn \\
m_b (m_b) & = & 4.22 \pm 0.09 \GeV. \nn \\
M_t ({\rm pole}) & = & 178.0 \pm 4.3 \GeV \nn \\
m_t(m_t) & = & 169 \pm 4 \GeV \nn
\eea
In our analysis, we will calculate our observables at the energy scale $M_Z$, except
for the top mass for which we will use $m_t(m_t)$.
Therefore we need to convert our data to values at $M_Z$.
Define the running parameters:
\bea
\eta_i \equiv \left\{
\ba{cl}
\f{m_i(M_Z)}{m_i(m_i)} & \; {\rm for} \;\; i = c,\ b \\
& \\
\f{m_i(M_Z)}{m_i(2 \GeV)} & \; {\rm for} \;\; i = u,\ d,\ s.
\ea
\right.
\eea
At two loops in QCD we find
\bea
\eta_c = 0.56, \hspace{7mm} \eta_b = 0.69, \\
\eta_u \; = \; \eta_d \; = \; \eta_s \; = \; 0.65.  \nn
\eea

The lepton masses at $M_Z$ from \cite{Fusaoka:1998vc} are
\bea
m_e & = & 0.48684727 \pm 0.00000014 \MeV \\
m_\mu & = & 102.75138 \pm 0.00033 \MeV \nn \\
m_\tau & = & 1746.7 \pm 0.3 \MeV. \nn
\eea

Also from \cite{Fusaoka:1998vc}, we take the gauge couplings at $M_Z$:
\bea
\alpha_1(M_Z) & = & 0.016829 \pm 0.000017 \\
\alpha_2(M_Z) & = & 0.033493 \pm 0.000059 \nn \\
\alpha_3(M_Z) & = & 0.118 \pm 0.003 \nn
\eea

We use the mass of the Z-boson to set the weak energy scale:~\cite{Hagiwara:fs}
\bea
M_Z & = & 91.1876 \pm 0.0021 \GeV
\eea
In our model, we do not calculate the effects of electro-weak symmetry breaking. We assume
that it happens properly and gives the correct weak scale masses. For this reason, we
use a calculated Higgs VEV at $M_Z$.
\bea
v & = & \f{M_Z}{\sqrt{\pi} \sqrt{\f{3}{5} \alpha_1(M_Z) + \alpha_2(M_Z)}} \\
& = & 246.41 \pm 0.17 \GeV \nn
\eea

Additional information is needed to calculate $\varepsilon_K$ in our model. We use the
CKM parametrization-independent formula
\bea
| \varepsilon_K | = \f{C_\varepsilon}{2} B_K
\left[ {\rm Im} \left( \f{(\Box_c)^2}{\Box_u} \right) S_c \sigma_c + {\rm Im} \left( \f{(\Box_t)^2}{\Box_u} \right) S_t \sigma_t
+ 2 \, {\rm Im} \left( \f{\Box_c \Box_t}{\Box_u} \right) S_{ct} \sigma_{ct} \right]
\eea
where
\bea
\Box_i & \equiv & V_{ud} V^*_{us} V_{is} V^*_{id}
\eea
and the various $S_i$ functions are Inami-Lim functions \cite{Inami:1980fz}. The
$\sigma_i$ are ${\cal O}(1)$ factors which we take from Battaglia \cite{Battaglia:2003in}
\bea
\sigma_c = 1.32 \;\;\;\;\; \sigma_t = 0.57 \;\;\;\;\; \sigma_{ct} = 0.47.
\eea
$C_\varepsilon$ is a ratio of well known low-energy observables. From Battaglia \cite{Battaglia:2003in},
this combination is
\bea
C_\varepsilon & = & \f{G_F^2 f_{K^+}^2 m_{K^0} M_W^2}{6 \pi^2 \sqrt{2} \Delta m_K }
= 3.837 \times 10^4.
\eea
We neglect the error associated with $C_\varepsilon$ since it is small and will
not affect the theoretical error in $\varepsilon_K$.
A large part of this theoretical error comes from $B_K$ for which we use \cite{Battaglia:2003in}.
\bea
B_K & = & 0.86 \pm 0.06 \pm 0.14 \\
& \simeq & 0.86 \pm 0.15 \nn \eea   where in the last line we add the statistical and systematic errors in
quadrature.

\section{Setup \label{sec:Setup} }

\subsection{Background \label{sec:Background} }
We consider a five dimensional supersymmetric SO(10) GUT compactified on an $S^1/(Z_2 \times Z'_2)$ orbifold
where $S^1$ is described by $y \in [0,2\pi R]$. The first orbifold, $Z_2$,
under which $y \rightarrow -y$, breaks 5D N=1 supersymmetry (4D N=2) to 4D N=1.
The other orbifold, $Z'_2$, under which $y \rightarrow -y + \pi R$,
breaks SO(10) down to the PS gauge group ${\rm SU(4)}_C \times {\rm SU(2)}_L \times {\rm SU(2)}_R$.
The fundamental domain of the $y$ direction is the line segment
$y \in [0,\pi R/2 ]$. SO(10) gauge symmetry is present everywhere except at the point $y=\pi R/2$,
which only has Pati-Salam gauge symmetry. Hence, we call the two inequivalent fixed points the
``SO(10)" ($y=0$) and ``Pati-Salam" branes ($y=\pi R/2$) where
each fixed point is a three-brane (3+1 dimensional spacetime). The Higgs mechanism on the PS brane completes
the breaking of the PS gauge symmetry to the SM gauge group.

The fields which live in the five dimensional space between the branes (known as the ``bulk") are even
or odd under the orbifold parities. We denote the parity of a given field under these orbifolds by two $\pm$ subscripts
on the field with the first corresponding to $Z_2$ and the second corresponding to $Z_2'$. This part of the setup,
including the orbifold structure, field parities, and supersymmetry and gauge symmetry breaking are based
on work done in \cite{Dermisek:2001hp} and \cite{Kim:2002im}. For more details, please see these references.

We wish to relate some of our fields by a family symmetry, D$_3$, under which the first and second family fields form a doublet.
Other fields within our model will be in various representations of D$_3$ which will affect the structure of the Yukawa matrices we generate.
We take the family symmetry to be independent of the orbifold symmetries.
A brief summary of the D$_3$ group is provided in Appendix \ref{sec:AppD_3},
where we give the information necessary to understand the family representations and couplings
used in our model. For more information, see the Appendix presented in \cite{Dermisek:1999vy}.

The 5D supersymmetric vector multiplet ${\cal V} = (A_M,\l_1,\l_2,\Sigma)$ contains a 4D vector multiplet $V = (A_\mu,\l_1)$
and a 4D chiral adjoint $\phi = ((\Sigma + i A_5)/\sqrt{2},\l_2)$.
For a generic hypermultiplet ${\cal H} = (h, \overline{h}, \psi, \overline{\psi})$ which breaks up into the 4D chiral multiplets $\Phi = (h, \psi)$ and
$\overline{\Phi} = (\overline{h}, \overline{\psi})$, we use Arkani-Hamed et.~al.~\cite{Arkani-Hamed:2001tb} for the 5D action:
\bea
S & \supset & \int d^4 x \, dy \left\{ \int d^4 \theta \left[ \overline{\Phi} e^V \overline{\Phi}^\dagger + \Phi^\dagger e^{-V} \Phi \right] \right. \\
&& \left. + \left[ \int d^2 \theta \ \overline{\Phi} ( m + \partial_y - \f{1}{\sqrt{2}} \phi ) \Phi + {\mbox h.c.} \right] \right\} \nn
\eea
The mass $m$ refers to the mass of the 5D fields, which we set to zero for all hypermultiplets. We take the field $\Sigma \; (\subset \phi)$
to get a VEV in the $X$ direction of SO(10). Because of the coupling in the action:
\bea
S & \supset & \int d^4 x \, dy \int d^2 \theta \ \overline{\Phi} (\partial_y - \f{1}{\sqrt{2}} \phi) \Phi \\
& \stackrel{\rm VEV}{\rightarrow} & \int d^4 x \, dy \int d^2 \theta \ \overline{\Phi} (\partial_y - m_X) \Phi \nn
\eea
this VEV will generate $X$-dependent masses $m_X$ for the bulk chiral fields coming from hypermultiplets in the theory.
The U(1)$_X$ of SO(10) is contained inside Pati-Salam, and the chiral adjoint for PS has $(--)$ parity. Therefore the
mass $m_X$ we have generated also has this odd-odd parity under the orbifold. For this reason we call it a kink mass.

The D-term for this theory is
\bea
D & = & -(\partial_y \Sigma + i \left[ \Sigma, A_5 \right] ),
\eea
hence a VEV of $\Sigma$ is potentially dangerous since it could create
a D-term VEV. To avoid this, we take $\langle \Sigma \rangle$ to be
flat in the bulk with discontinuities at the branes to be consistent with its $(--)$ parity.
These discontinuities generate D-term VEVs on the branes, but we can choose to add brane fields
($16_{\rm kink}$ on the SO(10) brane, $\chi^c_{\rm kink}$ and $\overline{\chi^c}_{\rm kink}$
on the PS brane)
with VEVs in the $\nu^c$ and $\overline{\nu^c}$ directions to cancel these
effects from $\partial_y \langle \Sigma \rangle$,
leaving us with a D-flat theory.\footnote{This technique is described in \cite{Arkani-Hamed:2001tb} and used in \cite{Kitano:2003cn}.}
More concretely,
if the sterile sneutrino component of
$16_{\rm kink}$ ($\nu^c_{\rm kink}$)
and the charge conjugate sterile sneutrino component of
$\overline{\chi^c}_{\rm kink}$
($\overline{\nu^c}_{\rm kink}$) get their VEVs,
the D-flat condition for the $U(1)_X$ subgroup
is given by
\bea
\partial_y \langle \Sigma \rangle & = & g^2 \left[
5 | \langle \nu^c_{\rm kink} \rangle |^2 \delta (y)
-5 | \langle \overline{\nu^c}_{\rm kink} \rangle |^2 \delta (y-\f{\pi R}{2})
\right].
\eea
As $\nu^c_{\rm kink}$ and $\overline{\nu^c}_{\rm kink}$ carry
only $U(1)_X$ charges, they do not appear in the D-term
of the standard model gauge group.
The D-flat condition is satisfied for
\bea
\langle \Sigma \rangle & = & \Sigma_0 \ve_{--} (y), \\
| \langle \nu^c_{\rm kink} \rangle |^2 & = &
| \langle \overline{\nu^c}_{\rm kink} \rangle |^2 = \frac{2 \Sigma_0}{5 g^2}.
\eea
$\ve_{-+}$ and $\ve_{--}$ are step functions on the orbifold.
Both of these will be used later in our analysis.
\bea
\varepsilon_{-+}(y) & \equiv & \left\{
\ba{ccccl}
-1 & \mbox{ for } & y & \in & [-\pi R, 0] \\
+1 & \mbox{ for } & y & \in & [0,\pi R]
\ea
\right. \label{eq:vedefs} \\
\varepsilon_{--}(y) & \equiv & \left\{
\ba{ccccl}
+1 & \mbox{ for } & y & \in & [-\pi R, -\f{\pi R}{2} ] \\
-1 & \mbox{ for } & y & \in & [-\f{\pi R}{2}, 0] \\
+1 & \mbox{ for } & y & \in & [0, \f{\pi R}{2} ] \\
-1 & \mbox{ for } & y & \in & [\f{\pi R}{2}, \pi R]
\ea
\right.
\eea

We choose to parametrize the kink mass
as follows:
\bea
m_X (y) & = & \ve_{--}(y) X \frac{\Sigma_0}{2}
= \ve_{--}(y) X \f{\zeta}{\pi R} \\
\zeta & \equiv & \frac{\Sigma_0 \pi R}{2}. \nn
\eea
$\zeta$ will turn out to be a useful dimensionless parameter
in later analysis.

If our $U(1)_X$-breaking fields were to get VEVs as large as the cutoff scale,
we could not keep SO(10) or Pati-Salam as our symmetries on the branes.
However, it turns out that $\zeta \sim 2$ is needed to fit the observed
physical quantities, and so the corresponding VEVs are
\bea
\langle \nu^c_{\rm kink} \rangle & \sim & \sqrt{\f{8}{5\pi R g^2}}
= \sqrt{\f{16}{5\pi^2 g_4^2 R^2}}
\sim \f{1}{R}. \nn
\eea
Therefore, all the VEVs necessary to give a kink mass are around
the compactification scale and the breaking effects are suppressed
at least by $M_c/M_* \sim 10^{-2}$ or $10^{-3}$ \cite{Kim:2002im}.
For the most part, the $U(1)_X$ breaking effects come from the kink profiles
of bulk fields which are calculable and are exponentially proportional
to the $U(1)_X$ charges.
It is a novel example of obtaining large (order one) symmetry-breaking
with a very small symmetry-breaking order parameter $\Sigma$.
We stress that the VEVs of $\Sigma$, $\nu^c_{\rm kink}$
and $\overline{\nu^c}_{\rm kink}$ do not spoil the symmetries on the branes.

On the other hand, on the Pati-Salam brane gauge coupling unification requires
$\chi^c$ and $\overline{\chi^c}$ to get VEVs of order the cutoff scale
($\overline{\chi^c}$ and $\chi^c$ are different from
$\overline{\chi^c}_{\rm kink}$ and $\chi^c_{\rm kink}$). This is a serious problem for our theory.
There are order one corrections
in the Kahler potential and Pati-Salam symmetry is badly broken in the
canonical basis:
\bea
K & = & (1 + C \f{\chi^{c\dagger}
\chi^c}{M_*^2} ) \psi^{\dagger} \psi
\eea
where $C$ is of order one, $\langle \chi^c \rangle = \langle \chi^{c\dagger} \rangle \sim M_*$,
and where $\psi$ is a generic Pati-Salam brane field.
This problem can be solved in a geometric way if there
is a sixth dimension along which $\chi^c$ and $\psi$ are
separated. If this sixth dimension has a length scale $R_2$ slightly larger than
the cutoff length scale, we can simultaneously retain the desired PS-brane Yukawa relations
and gauge coupling unification.
The 6D setup will be discussed in detail in section \ref{sec:6D}.
Until then, we choose to concentrate on the effective 5D theory below this 6D
scale. The 5D analysis of the Yukawa matrices will not be quantitatively affected by the addition of this extra dimension.

We now briefly summarize the gauge unification results given by Kim and Raby \cite{Kim:2002im}.
A 5D gauge theory is nonrenormalizable and gets large corrections at the cutoff. Corrections to gauge couplings,
however, will be the same for all couplings unified into the larger gauge group. These corrections will affect the absolute
values of the gauge couplings, but not the differences. Further, if the gauge symmetry is broken only
by orbifolding or by Higgs mechanism on the branes, the differences in the couplings will
have a logarithmic, calculable running.

The states which affect the differential running are the bulk vector multiplet ${\cal V}$ and the bulk Higgs
hypermultiplet ${\cal H}$.\footnote{ When considering differential running of the gauge couplings, a Higgs
hypermultiplet in the bulk is effectively the same as a 4D 10 of SO(10) with light Higgs MSSM doublets and heavy
Higgs triplets of mass $M_c$. This setup admits gauge coupling unification as shown by Kim and Raby
\cite{Kim:2002im}.  In particular, see the calculations leading to equation (3.13) of that paper. Effects from
brane Higgs doublets would be felt up to $M_*$ and would tend to inhibit unification since they drive the
couplings apart rather than together.  } The placement and number of complete matter multiplets does not affect
gauge coupling unification, since matter multiplets (16s of SO(10)) act equally across the three gauge couplings
and cannot affect the coupling differences. Those states in the theory outside of the MSSM have twisted orbifold
boundary conditions and so have masses at the compactification scale ($M_c$). For energies below $M_c$, the
theory is the MSSM. The effects of running between $M_c$ and $M_*$ (the cutoff scale), including the Kaluza-Klein
(KK) towers, are taken as threshold corrections at $M_c$. Without these threshold corrections, it is known that
the MSSM unifies around $M_G \sim 3 \times 10^{16} \GeV$ with a coupling of $\a_{\rm GUT} \sim 1/24$ and a
GUT-scale threshold correction for $\a_3$ of $\varepsilon_3 \sim -0.04$. Assuming unification in the orbifold
theory at the cutoff scale $M_*$ and that the PS breaking Higgs VEV is of order the cutoff scale, we can solve
for $M_*$ and $M_c$ in terms of the 4D GUT parameters $M_G$, $\a_{\rm GUT}$, and $\varepsilon_3$. This leads to
$M_* \sim 10^{17} \GeV$ and $M_c \sim 10^{14} \GeV$ \cite{Kim:2002im}.

The matter field locations are constrained by proton decay:

\begin{itemize}
\item Matter fields on the SO(10) brane

There are gauge bosons within SO(10) which mediate baryon (B) and lepton (L) number-violating interactions. All
of these are outside of the Standard Model and hence have masses of order $M_c$ or higher. After integrating out
these fields (and their KK modes), we get dimension six operators which violate B and L for any matter multiplets
on the SO(10) brane. These operators are suppressed by $1/M_c^2$. Given $M_c \sim 10^{14} \GeV$, current bounds
on proton decay rule out models which have these operators for the first and second families.  Thus only the
third family can reside on the SO(10) brane \cite{Murayama:2001ur}.\footnote{ With dimension six operators for
the third family, mixing between this family and the first two can induce proton decay. Assuming that the mixing
is of order $|V_{ub}|$ or $|V_{cb}|$ and that the gauge bosons have mass at $M_c$, naive calculations using
formulae in \cite{Murayama:2001ur} put the proton lifetime many orders of magnitude above current limits, since
this leads to an effective gauge boson mass of order $M_c/(V_{cb} V_{ub}) \approx 6 \times 10^{17}$ GeV.}
Dimension 5 proton decay operators vanish since the color triplet Higgs states obtain off-diagonal mass with
triplets in $\overline{10}$ and these states do not couple to matter.

\item Matter fields on the PS brane or in the bulk

Pati-Salam gauge symmetry does not relate the left-handed fields $\psi$ ($(4,2,1)$ in PS) to the
right-handed fields $\psi^c$ ($(\overline{4},1,2)$ in PS), and we do not get baryon number-violating dimension six operators
after integrating out the heavy gauge bosons. Therefore, any matter fields can be on the PS brane as long as the PS breaking
scale is not extremely low. This scale in our theory is $M_* \sim 10^{17} \GeV$, and proton decay is not a problem here.

In principle, we can consider higher dimensional operators
with derivative interactions $\partial_5 = \partial/ \partial y$.
Because the coefficients of the higher dimensional operators
are not determined from the theory we cannot calculate
the proton decay rate from these operators accurately.
However, we can get a bound that is consistent with our setup by assuming unknown coefficients to be
order one.
See Kim and Raby for more details \cite{Kim:2002im}.

\end{itemize}

Proton decay constrains the first and second families to reside either in the bulk
or on the PS brane, but does not constrain the location of the third family.
We choose to place the third family on the SO(10) brane.
Given this placement, the second family is forced to reside on the PS brane.
If it were in the bulk, volume suppression would give $m_\mu / m_\tau \sim 10^{-3}$, which is far
too small to fit the data. Our D$_3$ flavor symmetry places the first and
second families together into a doublet, so we place both on the PS brane.

Let us summarize the basic setup.
\begin{itemize}
\item Gauge symmetry :
SO(10) in the bulk and at $y=0$, Pati-Salam at $y=\pi R/2$.

\item Higgs fields come from 10 dimensional hypermultiplets in the bulk.

\item 3rd family matter fields are on the SO(10) brane.

\item 1st and 2nd family matter fields, a doublet under D$_3$, are on the PS brane.

\item Kink mass localizes the hypermultiplets through their $X$-dependence.

\end{itemize}

\subsection{Yukawa Matrix \label{sec:Yukawa} }

This section introduces the fields and superpotential of our model. Here we calculate the
Yukawa matrices associated with the massless fields corresponding to the Standard Model fermions.

First we introduce the matter content of the theory. In the bulk, we have two 5D N=1 hypermultiplets which
transform as $16$s under SO(10). These fields form a doublet $({\bold 2}_{\rm A})$ under the D$_3$ family symmetry.
In our Lagrangian, we show the D$_3$ doublet structure of these fields by a subscript $a$, with $a = 1,2$. For information on how
doublets and other D$_3$ objects couple, please see Appendix \ref{sec:AppD_3}.
We also have a hypermultiplet $10$ of SO(10) which is a singlet $({\bold 1}_{\rm A})$ under the family symmetry. These fields are listed in
Table \ref{t:bulkfields} with their parities, PS gauge symmetries, and family symmetries.
On the SO(10) brane we place a single $16$ of SO(10), invariant under D$_3$, and an SO(10) gauge singlet $\phi$ which
is a doublet under the family symmetry. These fields are listed in Table \ref{t:SO(10)fields}.
The PS brane has a more complicated set of fields. There are the fields $\psi$ and $\psi^c$, each doublets under D$_3$, which
transform as the two halves of an SO(10) $16$. $N^c$ and $\overline{N^c}$ are also family doublets and carry
charge under PS in order to allow mixing between $\psi^c$ and $N^c$. There are various other fields
$\widetilde{\phi}, \, A, \, A_{3R}, \, \omega^c,$ and $\overline{\omega^c}$ which will get vacuum expectation values. These
fields are listed in Table \ref{t:PSfields} with their PS charges and D$_3$ family symmetries.

\begin{table}
\caption{Bulk fields}
\label{t:bulkfields}
$$
\ba{lccc}
\mbox{Field} \;\; & \mbox{PS Symm} \;\;\;\; & {\rm D}_3 \mbox{ Symm} \\
\hline
16 =
\left(
\ba{c}
\psi_{--} \\
\psi^c_{-+}
\ea
\right) &
\left(
\ba{c}
(4,2,1) \\
(\overline{4}, 1 , 2)
\ea
\right) &
{\bf 2_A} \\
\overline{16} =
\left(
\ba{c}
\overline{\psi}_{++} \\
\overline{\psi^c}_{+-}
\ea
\right) &
\left(
\ba{c}
(\overline{4},2,1) \\
(4, 1, 2)
\ea
\right) &
{\bf 2_A} \\
10 =
\left(
\ba{c}
H_{++} \\
H^c_{+-}
\ea
\right) &
\left(
\ba{c}
(1,2,2) \\
(6,1,1)
\ea
\right) &
{\bf 1_A} \\
\overline{10} =
\left(
\ba{c}
\overline{H}_{--} \\
\overline{H^c}_{-+}
\ea
\right) &
\left(
\ba{c}
(1,2,2) \\
(6,1,1)
\ea
\right)
&
{\bf 1_A}
\ea
$$
\end{table}

\begin{table}
\caption{SO(10) Brane fields}
\label{t:SO(10)fields}
$$
\ba{lcc}
\mbox{Field} \;\; & \mbox{PS Symm} \;\;\;\; & {\rm D}_3 \mbox{ Symm} \\
\hline
16_3 =
\left(
\ba{c}
\psi_3 \\
\psi^c_3
\ea
\right)
&
\left(
\ba{c}
(4,2,1) \\
(\overline{4}, 1 , 2 )
\ea
\right)
&
{\bf 1_A} \\
\phi &
(1,1,1) &
{\bf 2_A} \\
\ea
$$
\end{table}

\begin{table}
\caption{PS Brane fields}
\label{t:PSfields}
$$
\ba{lcc}
\mbox{Field} \;\; & \mbox{PS Symm} \;\;\;\; & {\rm D}_3 \mbox{ Symm} \\
\hline
\psi &
(4,2,1) &
{\bf 2_A} \\
\psi^c &
(\overline{4},1,2) &
{\bf 2_A} \\
N^c &
(\overline{4},1,2) &
{\bf 2_A} \\
\overline{N^c} &
(4,1,2) &
{\bf 2_A} \\
\widetilde{\phi} &
(1,1,1) &
{\bf 2_A} \\
A &
(1,1,1) &
{\bf 1_B} \\
A_{3R} &
(1,1,3) &
{\bf 1_A} \\
\omega^c &
(\overline{4},1,2) &
{\bf 1_A} \\
\overline{\omega^c} &
(4,1,2) &
{\bf 1_A}
\ea
$$
\end{table}

We choose to break the superpotential into two pieces: $W = W_1 + W_2$.
The first contains terms leading to interactions between the matter and Higgs field ($H \equiv H_{++}$):
\bea
W_1 & = & \left[ \f{1}{2} \lambda_3 16_3 10 16_3 \right] \delta (y) \label{eq:W_1} \\
&& + \left[ \lambda_\alpha \psi_a H \psi^c_a A
+ \lambda_\beta \psi_a H N^c_a \widetilde{\phi}_a
+ 2 \lambda_\gamma \psi_a H A_{3R} \left( \psi^c_{-+} \right)_a \right] \delta (y-\f{\pi R}{2}) \nn
\eea
$W_2$ contains mass terms:
\bea
W_2 & = & \overline{16}_a (\partial_y - m_X) 16_a + \overline{10} (\partial_y - m_X) 10 \label{eq:W_2} \\
&& + \left[ 2 \sigma \phi_a \overline{16}_a 16_3 \right] \delta (y) \nn \\
&& + \left[ 2 \eta \left( \overline{\psi}_{++} \right)_a \chi_a + \overline{N^c}_a \left( a \psi^c_a + b_0
\omega^c \overline{\omega^c} N^c_a \right) \right] \delta (y-\f{\pi R}{2}) \nn \eea Sufficient factors of the
cutoff scale $M_*$ should be placed in the superpotential terms so that the couplings $\l_3$, $\l_\a$, $\l_\b$,
$\l_\g$, $\sigma$, $\eta$, $a$, and $b_0$ are dimensionless. We assume that at this stage our Lagrangian
possesses a CP-symmetry and that these couplings are all real. All CP-violation will come from spontaneous
symmetry breaking. The fields which get VEVs are $\phi_1$, $\phi_2$, $\widetilde{\phi}_2$, $A$, $A_{3R}$,
$\omega^c$, and $\overline{\omega^c}$. The first 4 break D$_3$ symmetry, while $\left< A_{3R} \right> = A^0_{3R}
T_{3R}$ breaks SU(2)$_R$ symmetry. All of these VEVs we take to be real. $\omega^c$ and $\overline{\omega^c}$ are
fields which get VEVs in the right-handed neutrino direction. These VEVs we take to be complex, and furthermore
take the combination to be in a particular direction: $\left< \omega^c \overline{\omega^c} \right> \propto c e^{i
\theta_1} + X_R e^{i \theta_2}$. More will be said later about this choice of direction. All of the above
mentioned VEV values are taken to be of order $M_*$ rather than $M_c$, but are allowed to lie enough below $M_*$
to be able to give the desired hierarchies in the Yukawa matrices. We also assume that the fields listed here,
which get VEVs,  obtain mass near the cutoff scale $M_*$. This lifts these fields high enough that they cannot
adversely affect the gauge coupling unification results we wish to preserve.

Our superpotential allows 9 extra U(1) symmetries which we take to be
symmetries of the theory in order to forbid problematic superpotential terms.\footnote{We can avoid unwanted
flavor changing neutral currents from the breaking of these extra symmetries by instead assuming only $Z_n$ subgroups of the U$(1)$s
sufficient to forbid the unwanted superpotential terms.}
We choose to parametrize these symmetries by the following 9 fields:
$H^c_{+-}$, $\psi_3$, $\psi^c_3$, $\psi_a$, $A$, $A_{3R}$, $\omega^c$, and $\overline{\omega^c}$. We choose each of these fields to have
an arbitrary charge under one and only one U(1). After specifying these charges, the charges of the remaining fields are determined
by the terms in the superpotential. In addition to these U(1)s, the superpotential can have $Z_2$ symmetries which are extensions
of the first orbifold $Z_2$. The transformations of the bulk fields under this symmetry have already been
given. To have such a symmetry, we can choose $A_{3R}$ odd and the remaining 7 independent fields ($H^c_{+-}$ is already determined) even under this
$Z_2$. The transformations of the remaining fields can then be determined from the superpotential terms. Another possible choice is $\psi_a$ odd with
the remaining fields even. It can be shown that it is not possible to similarly extend the second $Z_2$ orbifold to the brane terms of the
superpotential.

Our 5D bulk Higgs field $H(x^\mu,y)$ can be represented as a 4D massless state and a KK tower of 4D massive
states. In order to find the effective Yukawa matrices, we need to know the overlap between the 5D bulk field
$H(x^\mu,y)$ and the effective 4D massless Higgs $H^0(x^\mu)$.\footnote{When addressing the overlap between 5D
and 4D fields, to avoid confusion we will explicitly show the dependence of the fields on the spacetime
coordinates.} This overlap is \bea H(x^\mu,y) \supset e^{\f{X_H \z y}{\pi R}} n_H \rho \sqrt{M_*} \, H^0(x^\mu)
\label{eq:Higgswave} \eea where \bea
n_X & \equiv & \sqrt{\f{X \z}{e^{X \z}-1}} \label{eq:n_X} \\[2mm]
\rho & \equiv & \sqrt{\f{2 M_c}{\pi M_*}}. \nn
\eea
The kink mass dependence gives the massless Higgs field an exponential profile, localizing it to an end of the extra dimension.
We will use $n_X$, $n_H$, $n_L$, $n_R$, etc. to stand for a normalization as above with the particular value for the $X$ quantum number
substituted for the $X$ in equation (\ref{eq:n_X}). In the same manner, at times we will use $n_1$ or $n_{-3}$ to refer to these normalization
factors with the indicated $X$ value inserted.
We can see explicitly in equation (\ref{eq:Higgswave}) that the up- and down-Higgs wavefunctions
are localized towards opposite ends of the $y$ direction domain $\left[0,\f{\pi R}{2} \right]$ due to their opposite $X$ quantum numbers.

The light fields which will correspond to our observed three families of particles come from
various mixings of the fields in the superpotential. Consider the following terms contained in $W_2$:
\bea
W_2 \supset \left[ \overline{N^c}_a \left( a \psi^c_a + b_0 \omega^c \overline{\omega^c} N^c_a \right) \right] \delta (y-\f{\pi R}{2})
\eea
With the factors of $M_*$ explicit:
\bea
b_0 \left( \f{\omega^c \overline{\omega^c}}{M_*^2} \right) \stackrel{\rm VEV}{\rightarrow} b_0^\prime (c e^{i \theta_1} + X_R e^{i \theta_2}) \equiv b
\eea
One combination of $\psi^c_a$ and $N^c_a$ is massive while the other combination leads to a massless field. In terms
of $a$ and $b$, the overlap
between the massless field $\psi^{c \, 0}_a$ and the original fields is:
\bea
\psi^c_a & \supset & n^c \psi^{c \, 0}_a \\
N^c_a & \supset & -\f{a}{b} n^c \psi^{c \, 0}_a \nn \\
n^c & \equiv & \f{1}{\sqrt{1 + \f{a^2}{|b|^2}}} \nn
\eea
The light fields $\psi^{c \, 0}_a$ form a D$_3$ doublet and correspond to the 1st and 2nd family right-handed particles.
The factor of $\f{a}{b}$ in the overlap between $N^c_a$ and $\psi^{c \, 0}_a$ will lead to
$\f{1}{b} \propto \f{1}{c e^{i\theta_1} + X_R e^{i\theta_2}}$ in the
(2,2) elements of the Yukawa matrices. This factor is important in fitting inter-family fermion mass ratios and in
providing us with nontrivial phases in the Yukawa matrices.

Now we turn to a different subset of $W_2$:
\bea
W_2 & \supset & \overline{16}_a (\partial_y - m_X) 16_a \\
&& + \left[ 2 \sigma \phi_a \overline{16}_a 16_3 \right] \delta (y)
+ \left[ 2 \eta \left( \overline{\psi}_{++} \right)_a \chi_a \right] \delta(y - \f{\pi R}{2}) \nn
\eea
These mass terms lead to mixing between the brane fields $16_3 (x^\mu)$, $\chi_a (x^\mu)$ and the bulk fields $16_a (x^\mu,y)$.
The overlap between the resulting left-handed massless field $(\psi^0_3 (x^\mu))$ and those fields in the superpotential:
\bea
\psi_3 (x^\mu) & \supset & \widetilde{n}_L \psi^0_3 (x^\mu) \label{eq:3rdleft} \\
\left( \psi_{--} \right)_a (x^\mu,y) & \supset &
-\varepsilon_{--}(y) e^{\f{X_L \zeta y}{\pi R}} \left( \f{\phi_a}{M_*} \right) \sigma \sqrt{M_*} \, \widetilde{n}_L \psi^0_3 (x^\mu) \nn \\
\chi_a (x^\mu) & \supset & - e^{\f{X_L \zeta}{2}} \left( \f{\phi_a}{M_*} \right) \f{\sigma}{\eta} \widetilde{n}_L \psi^0_3 (x^\mu) \nn
\eea
The overlap in the right-handed fields is
\bea
\psi^c_3 (x^\mu) & \supset & \widetilde{n}_R \psi^{c \, 0}_3 (x^\mu) \\
\left( \psi^c_{-+} \right)_a (x^\mu,y) & \supset &
-\varepsilon_{-+}(y) e^{\f{X_R \zeta y}{\pi R}} \left( \f{\phi_a}{M_*} \right) \sigma \sqrt{M_*} \, \widetilde{n}_R \psi^{c \, 0}_3 (x^\mu). \nn
\eea
The definitions of $\ve_{-+}(y)$ and $\ve_{--}(y)$ have been given in equation (\ref{eq:vedefs}).
$\psi^0_3$ and $\psi^{c \, 0}_3$ correspond to the left- and right-handed 3rd family fields.
Other definitions follow:
\bea
\widetilde{n}_L & \equiv & \f{1}{\sqrt{1+r^2 \left[ \f{1}{n_L^2} + \f{\rho^2}{\eta^2} e^{X_L \zeta} \right]}} \\
\widetilde{n}_R & \equiv & \f{1}{\sqrt{1+r^2 \f{1}{n_R^2}}} \nn \\
r^2 & \equiv & \f{\sigma^2}{\rho^2} \left[ \left( \f{\phi_1}{M_*} \right)^2 + \left( \f{\phi_2}{M_*} \right)^2 \right] \nn
\eea
For more information on our treatment of these overlaps and normalizations, please see Appendix \ref{sec:AppWave}.

The left-handed 1st and 2nd family fields are equal to $\psi_a$, which do not mix with any
other fields. We replace all fields within $W_1$ with their massless components. This yields the $X$-dependent
Yukawa matrices for the massless fields with left-handed doublets on the left:
\bea
Y & = &
\left(
\ba{ccc}
0 & \a_0 \f{n^c}{L R} & \ve_0 \f{2 T_{3R}}{L} \\
- \a_0 \f{n^c}{L R} & \b_0 \f{b_0'}{b} \f{n^c}{L R} & \g_0 \f{2 T_{3R}}{L} \\
0 & 0 & 1
\ea
\right) \l.
\eea
Definitions follow:\footnote{We use the fact that for all Yukawa terms, $X_L + X_R + X_H = 0$.}
\bea
L & \equiv & e^{\f{X_L \z}{2}} \widetilde{n}_L \\
R & \equiv & e^{\f{X_R \z}{2}} \widetilde{n}_R \nn \\
\l & \equiv & \l_3 \rho \widetilde{n}_L \widetilde{n}_R n_H \nn \\
\a_0 & \equiv & \f{\l_\a}{\l_3} \f{\left< A \right>}{M_*} \nn \\
\b_0 & \equiv & - \f{\l_\b}{\l_3} \f{\left< \widetilde{\phi}_2 \right>}{M_*} \f{a}{b_0'} \nn \\
\g_0 & \equiv & - \f{\l_\g}{\l_3} \f{A^0_{3R}}{M_*} \f{ \left< \phi_1 \right>}{M_*} \sigma \nn \\
\ve_0 & \equiv & \g_0 \f{\phi_2}{\phi_1} \nn
\eea

The Yukawa matrices may be simplified by looking at the normalization constant $n^c$.
We assume that $a \ll |b|$, and so we can approximate $n^c \sim 1$. This is not incompatible with our definition
of $\b_0$, as we also expect $\b_0 \ll 1$.

An approximation may also be made within $\widetilde{n}_L$.
\bea
\widetilde{n}_L & \equiv & \f{1}{\sqrt{1+r^2 \left[ \f{1}{n_L^2} + \f{\rho^2}{\eta^2} e^{X_L \zeta} \right]}} \\
& \simeq & \f{1}{\sqrt{1+r^2 \f{1}{n_L^2}}} \nn
\eea
We assume that $\eta \gsim {\cal O}(1)$, and since $\rho \sim 1/40$ for reasonable values of the compactification
and cutoff scales derived from gauge coupling unification, $\f{\rho^2}{\eta^2} e^{X_L \zeta} \ll \f{1}{n_L^2}$ for
$-10 < \z < 10$, which easily encompasses the $\z$-range which has a chance of fitting the data.

It is convenient to reparametrize $r^2$ in terms of other variables. With the definition
\bea
\kappa & \equiv & \left( \f{\l_3}{\l_\g} \f{M_*}{A_{3 R}^0} \right)
\eea
$r^2$ may be rewritten as
\bea
r^2 & = & \f{\kappa^2}{\rho^2} (\g_0^2 + \ve_0^2).
\eea
With this redefinition the normalization constants $\widetilde{n}_L$ and $\widetilde{n}_R$ become
\bea
\widetilde{n}_X & = & \f{1}{\sqrt{1 + \f{\kappa^2}{\rho^2} \f{\g_0^2 + \ve_0^2}{n_X^2}}}
\eea
It is this definition for the $\widetilde{n}_X$ which we will use throughout the rest of this paper.

Under the approximation of $n^c \sim 1$, the Yukawa matrices simplify to:
\bea
Y & = &
\left(
\ba{ccc}
0 & \a_0 \f{1}{L R} & \ve_0 \f{2 T_{3R}}{L} \\
- \a_0 \f{1}{L R} & \b_0 \f{b_0'}{b} \f{1}{L R} & \g_0 \f{2 T_{3R}}{L} \\
0 & 0 & 1
\ea
\right) \l. \label{eq:Yukawa}
\eea
These are the Yukawa matrices which we will analyze in the next section.\footnote{We could have
included terms in our superpotential allowing (3,1) and (3,2) elements in the Yukawa matrices.
However, as long as such matrix elements are hierarchical, they cannot affect the theoretical
values of the low energy observables. Such terms are also not necessary in our theory (e.g., we do not
have a left-right symmetry which requires their presence) and so we have left them out
of our superpotential.}

The lowest order diagrams which contribute to the Yukawa matrices follow.
For each diagram we give the Yukawa element(s) to which it contributes.

\begin{center}
\SetPFont{Helvetica}{15}
\begin{picture}(300,280)(0,0)

\SetColor{Blue} \ArrowLine(75,260)(150,260) \Text(112.5,245)[cc]{$\psi_3$}

\SetColor{Red}  \ArrowLine(150,280)(150,260) \Text(140,280)[cc]{$H$}

\SetColor{Blue} \ArrowLine(225,260)(150,260) \Text(187.5,245)[cc]{$\psi_3^c$}

\SetColor{Black} \Text(340,260)[cc]{(3,3)}

\SetColor{Blue} \ArrowLine(0,180)(75,180) \Text(37.5,165)[cc]{$\psi_a$}

\SetColor{Red}  \ArrowLine(60,200)(75,180) \Text(40,200)[lc]{$H$}

\SetColor{Red}  \ArrowLine(90,200)(75,180) \Text(115,200)[rc]{$A_{3R}$}

\SetColor{Black} \ArrowLine(150,180)(75,180) \Text(112.5,165)[cc]{$ \left( \psi^c_{-+} \right)_a$}

\SetColor{Black} \Text(150,182)[cc]{X}

\SetColor{Black} \ArrowLine(150,180)(225,180) \Text(187.5,165)[cc]{$\left( \overline{\psi^c}_{+-} \right)_a$}

\SetColor{Red} \ArrowLine(225,200)(225,180) \Text(215,200)[cc]{$\phi_a$}

\SetColor{Blue} \ArrowLine(300,180)(225,180) \Text(262.5,165)[cc]{$\psi^c_3$}

\SetColor{Black} \Text(340,180)[cc]{(1,3) (2,3)}

\SetColor{Blue} \ArrowLine(0,100)(75,100) \Text(37.5,85)[cc]{$\psi_a$}

\SetColor{Red}  \ArrowLine(60,120)(75,100) \Text(50,120)[cc]{$H$}

\SetColor{Red}  \ArrowLine(90,120)(75,100) \Text(110,120)[rc]{$\widetilde{\phi}_a$}

\SetColor{Black} \ArrowLine(150,100)(75,100) \Text(112.5,85)[cc]{$N^c_a$}

\SetColor{Red} \ArrowLine(135,120)(150,100) \Text(125,120)[cc]{$\omega^c$}

\SetColor{Red} \ArrowLine(165,120)(150,100) \Text(185,120)[rc]{$\overline{\omega^c}$}

\SetColor{Black} \ArrowLine(150,100)(225,100) \Text(187.5,85)[cc]{$\overline{N^c}_a$}

\SetColor{Black} \Text(225,100)[cc]{X}

\SetColor{Blue} \ArrowLine(300,100)(225,100) \Text(262.5,85)[cc]{$\psi^c_a$}

\SetColor{Black} \Text(340,100)[cc]{(2,2)}

\SetColor{Blue} \ArrowLine(75,20)(150,20) \Text(112.5,5)[cc]{$\psi_a$}

\SetColor{Red}  \ArrowLine(135,40)(150,20) \Text(125,40)[cc]{$H$}

\SetColor{Red}  \ArrowLine(165,40)(150,20) \Text(185,40)[rc]{$A$}

\SetColor{Blue} \ArrowLine(225,20)(150,20) \Text(187.5,5)[cc]{$\psi^c_a$}

\SetColor{Black} \Text(340,20)[cc]{(1,2) (2,1)}

\end{picture}
\end{center}

\section{Analysis \label{sec:Analysis}}

We take two routes in the analysis of our model. In the first subsection we use analytic
methods to extract predictions from our theory. Relations between the Yukawa parameters and the
observables are given but are not solved due to the complexity of the equations.
A more precise numerical analysis, in which a full fit is achieved, is then given in the
following subsection.

\subsection{Analytic Fitting \label{sec:Analytic}}

The starting point for our analysis is the Yukawa matrix of equation (\ref{eq:Yukawa}).
In our model, this Yukawa matrix is defined at the compactification scale $M_c$.
We have taken into account the effects of integrating out the heavy fields but
have neglected the running effects on the Yukawa matrix elements between $M_*$
(where the superpotential of equations (\ref{eq:W_1}) and (\ref{eq:W_2}) is defined) and $M_c$.
We assume that these running effects are small enough to ignore at the level of a few percent.

To make the analysis easier, we define $\alpha_u, \alpha_d$, etc. below.
\bea
Y(X) & = & \left(
\ba{ccc}
0 & \a_0 \f{e^{X_H \z /2}}{\widetilde{n}_L \widetilde{n}_R} & \ve_0 2 T_{3 R} \f{e^{-X_L \z /2}}{\widetilde{n}_L} \\[2mm]
- \a_0 \f{e^{X_H \z /2}}{\widetilde{n}_L \widetilde{n}_R} & \ \f{\b_0}{c e^{i \theta_1} + X_R e^{i \theta_2}} \f{e^{X_H \z /2}}{\widetilde{n}_L \widetilde{n}_R} \
& \g_0 2 T_{3R} \f{e^{-X_L \z /2}}{\widetilde{n}_L} \\[2mm]
0 & 0 & 1
\ea
\right) \l_x \\
& \equiv &
\left(
\ba{ccc}
0 & \a_x & \ve_x \\
-\a_x & \b_x & \g_x \\
0 & 0 & 1
\ea
\right) \l_x \nn
\eea
After inputting the proper $X$ quantum numbers:
\bea
\ba{lll}
\a_u \equiv \a_0 e^{-\z} \f{1}{(\widetilde{n}_1)^2} & \;\;\;
\a_d \equiv \a_0 e^{\z} \f{1}{\widetilde{n}_1 \widetilde{n}_{-3}} & \;\;\;
\a_e \equiv \a_0 e^{\z} \f{1}{\widetilde{n}_1 \widetilde{n}_{-3}} \\[2mm]
\b_u \equiv \f{1}{c e^{i \theta_1} + e^{i \theta_2}} \b_0 e^{-\z} \f{1}{(\widetilde{n}_1)^2} & \;\;\;
\b_d \equiv \f{1}{c e^{i \theta_1} -3 e^{i \theta_2}} \b_0 e^{\z} \f{1}{\widetilde{n}_1 \widetilde{n}_{-3}} & \;\;\;
\b_e \equiv \f{1}{c e^{i \theta_1} + e^{i \theta_2}} \b_0 e^{\z} \f{1}{\widetilde{n}_1 \widetilde{n}_{-3}} \\[2mm]
\g_u \equiv -\g_0 e^{-\z /2} \f{1}{\widetilde{n}_1} & \;\;\;
\g_d \equiv \g_0 e^{-\z /2} \f{1}{\widetilde{n}_1} & \;\;\;
\g_e \equiv \g_0 e^{3 \z /2} \f{1}{\widetilde{n}_{-3}} \\[2mm]
\ve_u \equiv -\ve_0 e^{-\z /2} \f{1}{\widetilde{n}_1} & \;\;\;
\ve_d \equiv \ve_0 e^{-\z /2} \f{1}{\widetilde{n}_1} & \;\;\;
\ve_e \equiv \ve_0 e^{3 \z /2} \f{1}{\widetilde{n}_{-3}} \\[2mm]
\l_t & \;\;\;
\l_b = \lambda_t e^{-\z} \f{\widetilde{n}_{-3}}{\widetilde{n}_1} & \;\;\;
\l_\tau = \lambda_t e^{-\z} \f{\widetilde{n}_{-3}}{\widetilde{n}_1}
\ea \label{eq:Yelements}
\eea
and
\bea
\widetilde{n}_X & = & \f{1}{\sqrt{1 + \f{\kappa^2}{\rho^2} \f{\g_0^2 + \ve_0^2}{n_X^2}}}
\eea

There are eleven parameters associated with the fermion masses derived from the Yukawa matrices:
$\z$, $\a_0$, $\b_0$, $\g_0$, $\ve_0$, $c$, $\theta_1$, $\theta_2$, $\kappa$, $\l_t$, and $\tan \b$.
Because there are 13 independent observables in the quark and charged lepton sectors: 9 masses, 3 quark mixing angles, and the
CP-violating phase in the CKM matrix, we can expect two predictions in this model.

We assume that both of the quark Yukawa matrices are hierarchical \cite{Hall:1993ni}. As we explained within
\cite{Kim:2004ki} this allows us to use a simple set of rotations to diagonalize our quark mass matrices
and leads to a simple CKM matrix:
\bea
V_{CKM} & = & \left(
\ba{ccc}
1 & s^*_{12} + s^{U*}_{13} s_{23} & - s^{U*}_{12} s^*_{23} + s^*_{13} \\
-s_{12} - s^D_{13} s^*_{23} & 1 & s^*_{23} + s^U_{12} s^*_{13} \\
 s^D_{12} s_{23} -s_{13} & -s_{23} - s^{D*}_{12} s_{13} & 1
\ea
\right),
\eea
where
\bea
\ba{lll}
\vspace{2mm}
s^U_{12} \simeq \f{\alpha_u}{\beta_u}, \hspace{5mm} & s^D_{12} \simeq \f{\alpha_d}{\beta_d}, \hspace{5mm} & s_{12} \equiv s^D_{12} - s^U_{12}, \\
\vspace{2mm}
s^U_{13} \simeq \varepsilon_u, & s^D_{13} \simeq \varepsilon_d, & s_{13} \equiv s^D_{13} - s^U_{13}, \\
\vspace{2mm}
s^U_{23} \simeq \gamma_u, & s^D_{23} \simeq \gamma_d, & s_{23} \equiv s^D_{23} - s^U_{23}.
\ea
\eea
The eigenvalues of the diagonalized quark Yukawa matrices lead to the quark masses:
\bea
\ba{lll}
\vspace{2mm}
m_t \simeq \lambda_t \f{v_u}{\sqrt{2}} \hspace{5mm} & \f{m_c}{m_t} \simeq |\beta_u| \hspace{5mm} & \f{m_u}{m_c} \simeq \f{\alpha_u^2}{|\beta_u|^2} \\
\vspace{2mm}
m_b \simeq \lambda_b \f{v_d}{\sqrt{2}} & \f{m_s}{m_b} \simeq |\beta_d| & \f{m_d}{m_s} \simeq \f{\alpha_d^2}{|\beta_d|^2}
\ea
\eea

We will take the charged lepton Yukawa matrix to be lopsided to some extent, with $\g_e \sim {\cal O}(1)$ and the
rest of the matrix following a hierarchy.\footnote{The lopsided effect with order one $\g_e$ is
crucial to achieve correct b-$\tau$ unification
\cite{Barr:2002mw}.} This choice leads to the charged lepton masses:
\bea
\ba{lll}
m_\tau \simeq \lambda_\tau \f{v_d}{\sqrt{2}} \sqrt{1+\gamma_e^2} \hspace{5mm} & \f{m_\mu}{m_\tau} \simeq \f{|\beta_e|}{1+\gamma_e^2} \hspace{5mm} &
\f{m_e}{m_\mu} \simeq \f{\alpha_e^2}{|\beta_e|^2} \sqrt{1+\gamma_e^2}
\ea
\eea
We want to use the kink mass, here parametrized by $\z$, to localize the two Higgs wave functions to opposite branes. As stated before, if
$H_u$ is localized towards the SO(10) brane and $H_d$ is localized towards the PS brane, then there is a natural reason
why $m_t$ is larger than $m_b$ while $m_u$ is smaller than $m_d$. As can be seen from the $\a_x$ (equation (\ref{eq:Yelements})), which play a large
part in determining the masses of the first family fermions, if $\z > 0$ we have $m_u$ suppressed by $e^{-\z}$ and $m_d$ enhanced by $e^\z$.
Similarly, the third family masses have the opposite dependence as shown in the relation: $\l_b = \l_t e^{-\z} \f{\widetilde{n}_{-3}}{\widetilde{n}_1}$.
Looking at $\g_x$, we see that $\z > 0$ enhances $\g_e$ while suppressing $\g_u$ and $\g_d$, so it is not unreasonable to assume
that $\g_e \sim 1$ while $\g_u,\g_d \ll 1$.

With a few approximations within the CKM matrix the mixing angles and CP-violating angle $\b$ are:
\bea
|V_{us}| & \simeq & |s_{12}^* + s_{13}^{U \, *} s_{23}| \simeq |s_{12}|
\simeq \left| \f{\a_d}{\b_d} \right| \left| 1 - \f{\a_u \b_d}{\b_u \a_u} \right| \\[2mm]
|V_{cb}| & \simeq & |s_{23}^* + s_{12}^U s_{13}^*| \simeq |s_{23}| \simeq |\g_d - \g_u| \nn \\[2mm]
|V_{ub}| & \simeq & |s_{13}^* - s_{12}^{U \, *} s_{23}^*|
\simeq |\ve_d-\ve_u| \left| 1 - \f{\a_u}{\b_u} \f{(\g_d - \g_u)}{(\ve_d-\ve_u)} \right| \nn \\[2mm]
\b & \equiv & \arg \left( - \f{V_{cd} V_{cb}^*}{V_{td} V_{tb}^*} \right)
\simeq \arg \left(1 - \f{\a_u \b_d}{\b_u \a_d} \right) - \arg \left(1 - \f{\b_d}{\a_d} \f{(\ve_d-\ve_u)}{(\g_d-\g_u)} \right) \nn
\eea

With hierarchical structure in the up and down quark matrices it is the case that
$|V_{cb}| \sim \left| \f{(2,3)_D}{(3,3)_D} - \f{(2,3)_U}{(3,3)_U}  \right|$. Here this is reflected in the
relation $|V_{cb}| \simeq |\g_d - \g_u|$.
To have a nonzero $V_{cb}$, we need to have some dependence on right-handed quantum numbers
within $\g_x$. We have found that in our setup, regardless of the placement of fields, the kink mass effects (which come with $\z X$) can
only bring $X_L$ into the Yukawa element ratio $\f{(2,3)}{(3,3)}$. For this reason
we have added the field $A_{3R}$ to the superpotential,
and it is the VEV of this field (proportional to $T_{3R}$) which gives us the necessary difference between $\g_d$ and $\g_u$ to
generate a nonzero $V_{cb}$.

The structure of the $(2,2)$ Yukawa elements, proportional to $\f{1}{c e^{i \theta_1} + X_R e^{i \theta_2}}$, helps us to fit several
observables. The $X_R$ dependence gives different phases between the up and down elements, both of which contribute to
the CKM matrix CP-violating phase $\b$. The phase of $\b_u$ is also important in fitting the size of $|V_{ub}|$. In addition,
the magnitude of these $(2,2)$ elements helps us to fit the first to second and second to third family mass ratios.

We will now show the predictions of this model in the form of
relations between observables at the compactification scale.
Consider the fermion mass combinations $\left( \f{m_\tau}{m_b}
\right)$ and $\left( \f{m_u}{m_c} \right) / \left( \f{m_e}{m_\mu}
\right)$:
\bea
\left( \f{m_\tau}{m_b} \right) (M_c) & \simeq & \f{\l_\tau}{\l_b} \sqrt{1+\g_e^2} = \sqrt{1+\g_e^2} \\
\left( \f{m_u}{m_c} \right) / \left( \f{m_e}{m_\mu} \right) (M_c) & \simeq & \f{\a_u^2}{\a_e^2} \f{|\b_e|^2}{|\b_u|^2} \f{1}{\sqrt{1+\g_e^2}}
= \f{1}{\sqrt{1+\g_e^2}} \nn
\eea
One prediction of this model is then
\bea
\left( \f{m_u}{m_c} \right) / \left( \f{m_e}{m_\mu} \right) (M_c) & \simeq & \left( \f{m_b}{m_\tau} \right) (M_c).
\eea
Consider also the determinants of the down quark and charged lepton mass matrices.
\bea
m_d m_s m_b (M_c) & = & |\det Y_d| \left( \f{v_d}{\sqrt{2}} \right)^3 = \a_d^2 \l_b \left( \f{v_d}{\sqrt{2}} \right)^3 \\
m_e m_\mu m_\tau (M_c) & = & |\det Y_e| \left( \f{v_d}{\sqrt{2}} \right)^3 = \a_e^2 \l_\tau \left( \f{v_d}{\sqrt{2}} \right)^3 \nn
\eea
Because $\a_d = \a_e$ and $\l_b = \l_\tau$, we have the (exact) prediction:
\bea
m_d m_s m_b (M_c) & = & m_e m_\mu m_\tau (M_c)
\eea
These two predictions hold regardless of the actual values taken on by the Yukawa parameters.

As for the Yukawa parameter values themselves, these must be determined from the eleven remaining independent observables, if possible.
The dependence of these observables on the parameters is complicated enough that we find it impossible
to make a fit nonnumerically. Actual fit values for the Yukawa parameters must wait until the numerical analysis in Section \ref{sec:Numerical}.

Our predictions are relations at the compactification scale, while our data has been taken at the weak scale. In order to determine
the extent to which our predictions are reasonable, we need to estimate the renormalization effects on the masses
between the two scales. In this analysis, we choose to diagonalize the Yukawa
matrices at the compactification scale, and use the RG formalism of Barger, Berger, and Ohmann \cite{Barger:1992pk} to relate the observables
at $M_Z$ (or at $m_t$ in the case of the top quark) to their values at $M_c$ with simple scaling relations.
\bea
M_u^{\rm diag} (M_c) = \hspace{110mm} \nn \\[2mm]
\left(
\ba{ccc}
S_u(M_Z, M_c) m_u(M_Z) & 0 & 0 \\
0 & S_u(M_Z, M_c) m_c(M_Z) & 0 \\
0 & 0 & S_t(m_t, M_c) m_t(m_t)
\ea
\right) \\[3mm]
M_d^{\rm diag} (M_c) = \hspace{110mm} \nn \\[2mm]
\left(
\ba{ccc}
S_d(M_Z, M_c) m_d(M_Z) & 0 & 0 \\
0 & S_d(M_Z, M_c) m_s(M_Z) & 0 \\
0 & 0 & S_b(M_Z, M_c) m_b(M_Z)
\ea
\right) \nn \\[2mm]
M_e^{\rm diag} (M_c) = \hspace{110mm} \nn \\
\left(
\ba{ccc}
S_e(M_Z, M_c) m_e(M_Z) & 0 & 0 \\
0 & S_e(M_Z, M_c) m_\mu(M_Z) & 0 \\
0 & 0 & S_\tau(M_Z, M_c) m_\tau(M_Z)
\ea
\right) \nn \\[2mm]
|V|^2 (M_c) = \hspace{112mm} \nn \\
\left(
\ba{ccc}
|V_{ud}|^2(M_Z) & |V_{us}|^2(M_Z) & S(M_Z, M_c) |V_{ub}|^2(M_Z) \\
|V_{cd}|^2(M_Z) & |V_{cs}|^2(M_Z) & S(M_Z, M_c) |V_{cb}|^2(M_Z) \\
S(M_Z, M_c) |V_{td}|^2(M_Z) & S(M_Z, M_c) |V_{ts}|^2(M_Z) & |V_{tb}|^2(M_Z) \\
\ea
\right) \nn
\eea
The scaling factors $S$ can be found in \cite{Barger:1992pk}. In deriving these scaling factors, the authors have
used 2-loop running and have included running effects from the gauge couplings and the third family Yukawa couplings.
We make a further approximation in that we keep only 1-loop effects and neglect all running effects from the
Yukawa sector except for the top Yukawa coupling. With these approximations
\bea
S_u(M_Z, M_c) & \simeq & y_t^{-3}(m_t, M_c) G_u(M_Z, M_c) \label{eq:S_x} \\
S_t(m_t, M_c) & \simeq & y_t^{-6}(m_t, M_c) G_u(m_t, M_c) \nn \\
S_d(M_Z, M_c) & \simeq & G_d(M_Z, M_c) \nn \\
S_b(M_Z, M_c) & \simeq & y_t^{-1}(m_t, M_c) G_d(M_Z, M_c) \nn \\
S_e(M_Z, M_c) & \simeq & G_e(M_Z, M_c) \nn \\
S_\tau(M_Z, M_c) & \simeq & G_e(M_Z, M_c) \nn \\
S(M_Z, M_c) & \simeq & y_t^2(m_t, M_c) \nn
\eea
with
\bea
y_t(m_t, M_c) & \equiv & \exp \[ -\f{1}{16 \pi^2} \int_{m_t}^{M_c} \lambda_t^2(\mu) \, {\rm d} \! \( \log \mu \) \] \\
G_x(M_Z, M_c) & \equiv & \exp \[ -\f{1}{16 \pi^2} \int_{M_Z}^{M_c} \sum_i c^x_i g^2_i (\mu) \, {\rm d} \! \( \log \mu \) \] \nn \\
& = & \prod_i \left( \f{\alpha_i(M_Z)}{\alpha_i(M_c)} \right)_.^{ \f{c^x_i}{2 B_i} }
\eea
The $c^x$ and $B$ govern the MSSM 1-loop running effects due to the gauge couplings:
\bea
c^u & = & \left( \f{13}{15}, \, 3, \, \f{16}{3} \right) \\
c^d & = & \left( \f{7}{15}, \, 3, \, \f{16}{3} \right) \nn \\
c^e & = & \left( \f{9}{5}, \, 3, \, 0 \right) \nn \\
B & = & \left( \f{33}{5}, \, 1, \, -3 \right) \nn
\eea

We estimate that the neglect of the running due to $\lambda_b$ and $\lambda_\tau$ introduces 5\% error in the
down and charged lepton sectors and around $1\%$ error in the up sector.
We further estimate the neglect of 2-loop gauge running
to be $\sim 1\%$ error in $G_u$ and $G_d$, and $\sim 4\%$ error in $G_e$.

Because we use MSSM running from $M_c$ down to $M_Z$, errors are introduced when we include particles in the
running below their mass scales. These effects can be taken into account by weak-scale threshold corrections.
However, in our analysis we do not keep track of the supersymmetric particle masses and so these corrections
cannot be calculated. We can, nevertheless, estimate these effects. The SUSY thresholds for all of the down
sector quarks were calculated in \cite{Blazek:1995nv}. There are $\tan \b$-enhanced diagrams which contribute to
$m_b$, $m_s$ and $m_d$.  Although the Higgsino contribution to the $d, \; s$ threshold corrections are
Yukawa-suppressed, the dominant correction comes from gluino and Wino loop diagrams which affect all three of
these quarks equally (up to some small differences due to unequal squark masses). We therefore make the
simplifying approximation that the SUSY threshold corrections for all of the down quarks are the same. The same
approximation can be made separately for the up and charged lepton sectors where the differences in thresholds
among the same particle type arise only from squark mass differences, and are small.

By assuming some range of SUSY parameters, Pierce et. al.~\cite{Pierce:1996zz} have estimated the SUSY
threshold corrections to the 3rd family masses.
Assuming a value of $\tan \b \sim 30$ and $\mu > 0$,
we estimate from plots in \cite{Pierce:1996zz} the SUSY threshold effects
(at $m_t$ for top and $M_Z$ for bottom and tau) in terms of \% shifts and errors.
The shifts for these particles, and by approximation the shifts for the 1st and 2nd family particles, are
\bea
\ba{lll}
m_u, m_c, m_t \hspace{5mm} & \;\: (+ 2.0 \pm 3) \% \hspace{5mm} & F_t \equiv 1.02 \\
m_d, m_s, m_b & (- 10.0 \pm 10) \% & F_b \equiv 0.90 \\
m_e, m_\mu, m_\tau & \;\: (+1.5 \pm 2) \% & F_\tau \equiv 1.015. \ea \label{eq:FermionFs} \eea The factors $F_x$
will be used to keep track of these corrections.

SUSY threshold corrections to the CKM matrix elements were also calculated in \cite{Blazek:1995nv}. As those
authors state, to a good approximation the only CKM elements which shift their sizes are $|V_{ub}|$, $|V_{cb}|$,
$|V_{td}|$, and $|V_{ts}|$, and these shifts are the opposite of the chargino-induced shift in $m_b$. In
addition, $J$ shifts approximately twice the amount of the CKM elements. The contribution from charginos to the
$m_b$ shift is plotted separately in Pierce et. al.~\cite{Pierce:1996zz}, and we estimate it to be $5 \pm 10 \%$.
For those CKM observables which have large SUSY threshold corrections: \bea \ba{lll}
|V_{ub}|, |V_{cb}|, |V_{td}|, |V_{ts}| \hspace{5mm} & \;\: (-5.0 \pm 10) \% \hspace{5mm} & F_V \equiv 0.95 \\
J & (-10.0 \pm 20) \% & \, F_J \equiv 0.90 . \label{eq:CKMFs}
\ea
\eea
As explained in \cite{Blazek:1995nv}, each side of the
unitarity triangle has one element which has a large SUSY
threshold. This leads to the threshold for $J$, and also
implies that the angles of the triangle are unaffected.
Thus $\sin 2 \b$ does not get a correction. We assume that
the correction to $|\ve_K|$ is approximately the same size as the correction for $J$.

The mutual dependence of $y_t(m_t, M_c)$ and $\lambda_t$ means that there is no simple analytic
way of integrating to find $y_t(m_t, M_c)$. We choose, therefore, to fit $m_t(m_t)$ by adjusting
$\lambda_t(M_c)$ and using the 1-loop equations with $\lambda_t$ and $g_i$ only to run down to $m_t$.
The 1-loop RG equation is:
\bea
\f{d \lambda_t}{d t} & \simeq & \f{\lambda_t}{16 \pi^2} \left[ 6 \lambda_t^2 - \sum_i c^u_i g_i^2 \right]
\eea
with $t \equiv \log \mu$ where $\mu$ is the energy scale.
The relation between the experimental value of $m_t(m_t)$ and $\lambda_t(m_t)$:
\bea
m_t(m_t) & = & \lambda_t(m_t) F_t \f{v}{\sqrt{2}} \sin \beta \simeq \lambda_t(m_t) F_t \f{v}{\sqrt{2}}
\eea
We have already assumed $\tan \b \sim 30$, which means that $\sin \b$ is negligible, and we have included the thresholds
for $m_t(m_t)$ in the scale factor $F_t$.

The PS-breaking VEVs of $\chi^c$ and $\overline{\chi^c}$ are of order the cutoff scale. As in \cite{Kim:2002im},
we choose to parameterize these VEVs with a dimensionless parameter $\zeta_{\rm brane}$:
\bea
\langle \chi^c \rangle = \langle \overline{\chi^c} \rangle \equiv \sqrt{ \f{4 M_*}{\pi g_5^2 \zeta_{\rm brane} } }
\eea
Naive dimensional analysis leads to a value for $\zeta_{\rm brane}$:
\bea
\zeta_{\rm brane} = 0.27
\eea
Using this value and using
the assumption of 5D gauge unification as in \cite{Kim:2002im} with the 4D GUT scale inputs
\bea
M_G & = & (2.5 \pm 0.5) \times 10^{16} \GeV \label{eq:GUTparams} \\
\alpha_{\rm GUT} & = & 1/(24 \pm 1) \nn \\
\varepsilon_3 & = & -0.035 \pm 0.005 \nn
\eea
leads to knowledge of $M_*, \ M_c,$ and $\alpha_i(M_c)$:
\bea
M_* & = & 2.3 \times 10^{17} \GeV \\
M_c & = & 2.4 \times 10^{14} \GeV \nn \\
\a_1(M_c) & = & 0.035 \nn \\
\a_2(M_c) & = & 0.040 \nn \\
\a_3(M_c) & = & 0.044 \nn
\eea
The uncertainties in the
4D GUT scale parameters (especially $\alpha_{\rm GUT}$) introduce $\pm 9\%$ errors
into $G_u$ and $G_d$ and $\pm 2 \%$ error into $G_e$ and $y_t$. The errors listed here on $G_u$ and
$G_d$ have a high correlation and can be mostly neglected when considering ratios.

Considering all the sources of errors discussed above, we choose to parametrize our theoretical errors as $\pm
9\%$ $G_u$ and $G_d$ (with the errors highly correlated), $\pm 4.5\%$ on $G_e$, and $\pm 2\%$ on $y_t$. In
addition, in our expressions, the masses themselves should have extra theoretical errors from both threshold
corrections and the neglect of the contribution to Yukawa running due to $\lambda_b$ and $\lambda_\tau$. We
choose the up-type quarks to have $\pm 3\%$ error, down-type $\pm 11\%$ error, and the charged leptons $\pm 5\%$.
Each of these errors is correlated to some degree within each particle type.

Starting from the $\alpha_i(M_c)$ and using 1-loop running, we find the gauge coupling
scale factors to be
\bea
G_u(M_Z, M_c) & = & 0.33 \pm 0.03 \\
G_d(M_Z, M_c) & = & 0.33 \pm 0.03 \nn \\
G_e(M_Z, M_c) & = & 0.70 \pm 0.03. \nn
\eea
Fitting the central value of $m_t(m_t) = 169 \pm 4 \GeV$, allowing for the $m_t$ 1$\sigma$ range, and including
the $3\%$ threshold error on the top quark leads to
\bea
\lambda_t(M_c) & = & 0.59 \pm 0.12  \\
y_t(m_t, M_c) & = & 0.903 \pm 0.018. \nn
\eea
We have included the $2 \%$ theoretical error on $y_t$.

We are now in a position to analyze our predictions. First, consider
\bea
\left( \f{m_u}{m_c} \right) / \left( \f{m_e}{m_\mu} \right) (M_c) & \simeq & \left( \f{m_b}{m_\tau} \right) (M_c).
\eea
We choose to make a prediction for $\left( \f{m_u}{m_c} \right) (M_Z)$. The prediction and corresponding experimental
values are:
\bea
\left( \f{m_u}{m_c} \right)_{\rm th} (M_Z)
& \simeq & \left[ \left( \f{m_e}{m_\mu} \right) \left( \f{m_b}{m_\tau} \f{F_b}{F_\tau} \f{G_d }{G_e} \f{1}{y_t} \right) \right](M_Z) \\[2mm]
& \simeq & 0.0037 \pm 0.0006 \nn \\
\left( \f{m_u}{m_c} \right)_{\rm exp} (M_Z) & = & 0.0023 \pm 0.0010 \nn
\eea
Our prediction falls outside of the experimental bounds and leads to a 1.2$\sigma$ discrepancy with the data. Our model
favors a larger value for $\left( \f{m_u}{m_c} \right)$, indicating possibly a larger $m_u$ and/or smaller $m_c$ than currently measured.

Consider the second prediction:
\bea
m_d m_s m_b (M_c) & = & m_e m_\mu m_\tau (M_c)
\eea
After using the scale factors to get down to $M_Z$, this becomes
\bea
( m_d m_s m_b )_{\rm th} (M_Z) & \simeq & \left[ (m_e m_\mu m_\tau) \f{F_\tau^3}{F_b^3} \f{G_e^3}{G_d^3} y_t \right] (M_Z) \\[2mm]
& \simeq & (10.7 \pm 5.0) \times 10^5 \MeV^3 \nn \\[2mm]
( m_d m_s m_b )_{\rm exp} (M_Z) & = & (5.2 \pm 2.6) \times 10^5 \MeV^3. \nn
\eea
The uncertainty of $50 \%$ in the predicted value comes mostly from the $9 \%$ errors in the
gauge running scale factors $G_e$ and $G_d$ and the $11\%$ error associated with the
down-type masses. The experimental error is dominated by the uncertainty in
$m_s$ (and hence $m_d$).
The discrepancy in the two values is about 1$\sigma$, and we predict
a slightly larger scale for the down-type masses than measured.

\subsection{Numerical Fitting \label{sec:Numerical} }

Within this section we apply numerical methods to fit our model to the data. By using automated techniques,
we can find all of the Yukawa parameters which fit the data, including those parameters which were difficult
to determine analytically. We can also easily include 2-loop gauge running and (1-loop)
running due to the whole Yukawa matrices, not just $\l_t$.

Our fitting procedure starts from the Yukawa matrix in equation (\ref{eq:Yukawa}) taken at the compactification scale. We
determine the gauge couplings at that scale assuming unification and some usual values for the
unification parameters as listed in equations (\ref{eq:GUTparams}).
Using 2-loop gauge and 1-loop Yukawa MSSM running, the Yukawa matrices are run from the compactification scale down to
$M_Z$ and diagonalized to find the fermion masses and quark mixing angles. On the way to $M_Z$,
the top running mass $m_t(m_t)$ and the running due to $\l_t$ below $m_t$ ($y_t(M_Z, m_t)$) are determined.
The observables at $M_Z$ are shifted by SUSY threshold effects, listed in equations (\ref{eq:FermionFs}) and (\ref{eq:CKMFs}).
To counteract the inclusion of the incorrect top running between $M_Z$ and $m_t$, we
multiply the observables at $M_Z$ by appropriate powers of $y_t(M_Z, m_t)$. The correct
powers are determined from the scale factors listed in the previous section equations (\ref{eq:S_x}).
The observables are then compared to data in a $\chi^2$ function, which we minimize by altering the Yukawa input parameters.

\begin{table}
\caption{Observables used in the $\chi^2$ analysis. The theoretical errors are combinations of estimates of weak threshold
effects, GUT parameter uncertainties, and in the case of $\varepsilon_K$ an additional theoretical uncertainty from $B_K$.
All observables are at the $M_Z$ energy scale, except for the top mass, which is at $m_t$.}
\label{t:ChiObservables}
$$
\ba{lllll}
\mbox{Observable} \;\; & \mbox{Exp. Error \%} \;\;\;\; & \mbox{Th. Error \%} \;\;\;\; & \mbox{Total Error \%} \;\;\;\; & \mbox{Value Used} \\
\hline
Q & \;\: 3.5 & \;\: 6.4 & \;\: 7.3 & 22.7 \pm 1.7 \\
m_c & 23.1 & \;\: 3.6 & 23.4 & 0.73 \pm 0.17 \GeV \\
m_t(m_t) & \;\: 2.4 & \;\: 3.2 & \;\: 4.0 & 169 \pm 7 \GeV \\
\hline
\f{m_s}{m_d} & \;\: 8.5 & \;\: 2.5 & \;\: 8.8 & 18.9 \pm 1.7 \\
\f{m_s}{m_b} & 24.8 & \;\: 2.2 & 24.9 & 0.0199 \pm 0.0049 \\
m_b & \;\: 2.1 & 12.2 & 12.4 & 2.91 \pm 0.36 \GeV \\
\hline
\f{m_e}{m_\mu} & \;\: 0.0003 & \;\: 2.0 & \;\: 2.0 & 0.004738 \pm 0.000095 \\
\f{m_\mu}{m_\tau} & \;\: 0.02 & \;\: 2.0 & \;\: 2.0 & 0.0588 \pm 0.0012 \\
m_\tau & \;\: 0.02 & \;\: 2.0 & \;\: 2.0 & 1.747 \pm 0.035 \GeV \\
\hline
|V_{us}| & \;\: 1.6 & \;\: 0.0 & \;\: 1.6 & 0.2240 \pm 0.0036 \\
|V_{cb}| & \;\: 1.9 & 10.0 & 10.2 & 0.0415 \pm 0.0042 \\
\left| V_{ub}/V_{cb} \right| & \;\: 9.3 & \;\: 0.0 & \;\: 9.3 & 0.086 \pm 0.008 \\
|V_{td}| & \;\: 9.8 & 10.0 & 14.0 & 0.0082 \pm 0.0011 \\
\hline
\sin 2 \beta & \;\: 6.5 & \;\: 0.0 & \;\: 6.5 & 0.739 \pm 0.048 \\
J \times 10^5 & 10.0 & 20.0 & 22.4 & 3.0 \pm 0.7 \\
| \varepsilon_K | & \;\: 0.7 & 26.6 & 26.6 & 0.00228 \pm 0.00061 \\
\hline
\ea
$$
\end{table}

The sources of error in this section are the following. First, as in the analytic section, the threshold effects
on the observables at the weak scale due to the SUSY particle spectrum are not calculated. We estimate these
effects in the same way as in the previous section in equations (\ref{eq:FermionFs}) and (\ref{eq:CKMFs}).
Second, the uncertainties in our canonical 4D GUT parameters still introduce theoretical errors in our low energy
data. By altering these inputs and studying their effects on the calculated observables numerically, we estimate
the theoretical errors on our observables from these effects. These estimates are in general a bit less than the
errors assigned in the analytic section because some correlations are automatically included numerically which
were not included before. Third, we have the usual experimental errors. We choose to combine errors in quadrature
and assign a combined theoretical and experimental error to each observable. These observables and errors are
listed in Table \ref{t:ChiObservables}. Most theoretical errors cancel in the same-family mass ratios, up to a
few percent.  Hence, we have assigned a minimum $2\%$ error on these ratios due to threshold effects. This is
especially important for the lepton ratios, whose experimental errors are negligible.

Taken from Martin and Vaughn \cite{Martin:1993zk}, we use 2-loop gauge and 1-loop Yukawa renormalization group equations.
As explained before, our boundary conditions are at $M_c$:
Given a set of choices of the Yukawa parameters, we have the 3 $\times$ 3 complex Yukawa matrices at $M_c$;
with the assumption of gauge unification and with the 4D GUT parameters (equation (\ref{eq:GUTparams})), we have $\alpha_i(M_c)$.
Running due to the neutrino sector of the theory is neglected. We fit the 11 Yukawa parameters to the 16 observables listed
in Table \ref{t:ChiObservables} by minimizing a $\chi^2$ function
\bea
\chi^2 & = & \sum_{i=1}^N \f{(X^i_{\rm Calc}-X^i_{\rm Exp})^2}{(\sigma_X^i)^2}.
\eea

\begin{table}
\caption{Observables, target values, best fit values, and $\chi^2$ contributions.
All observables are at the $M_Z$ energy scale, except for the top mass, which is at $m_t$.}
\label{t:ChiResults}
$$
\ba{llll}
\mbox{Observable} \hspace{5mm} & \mbox{Target Value} \hspace{15mm} & \mbox{Fit Value} \hspace{10mm} & \chi^2 \mbox{ Contribution} \\
\hline
Q & 22.7 \pm 1.7 & 23.5 & 0.21 \\
m_c & 0.73 \pm 0.17 \GeV & 0.59 \GeV & 0.67 \\
m_t(m_t) & 169 \pm 7 \GeV & 167 \GeV & 0.11 \\
\hline
\f{m_s}{m_d} & 18.9 \pm 1.7 & 17.0 & 1.28 \\
\f{m_s}{m_b} & 0.0199 \pm 0.0049 & 0.0238 & 0.64 \\
m_b & 2.91 \pm 0.36 \GeV & 2.46 \GeV & 1.57 \\
\hline
\f{m_e}{m_\mu} & 0.004738 \pm 0.000095 & 0.004729 & 0.01 \\
\f{m_\mu}{m_\tau} & 0.0588 \pm 0.0012 & 0.0588 & 0.00 \\
m_\tau & 1.747 \pm 0.035 \GeV & 1.757 \GeV & 0.08 \\
\hline
|V_{us}| & 0.2240 \pm 0.0036 & 0.2237 & 0.01 \\
|V_{cb}| & 0.0415 \pm 0.0042 & 0.0412 & 0.00 \\
\left| V_{ub}/V_{cb} \right| & 0.086 \pm 0.008 & 0.090 & 0.21 \\
|V_{td}| & 0.0082 \pm 0.0011 & 0.0084 & 0.03 \\
\hline
\sin 2 \beta & 0.739 \pm 0.048 & 0.720 & 0.15 \\
J \times 10^5 & 3.0 \pm 0.7 & 3.0 & 0.00 \\
| \varepsilon_K | & 0.00228 \pm 0.00061 & 0.00210 & 0.09 \\
\hline
& & \mbox{Total:} & 5.06 \\
\ea
$$
\end{table}

\begin{table}
\caption{Minimum $\chi^2$ fit Yukawa parameters.}
\label{t:FitParams}
$$
\ba{lll}
\mbox{Parameter} \hspace{5mm} & \mbox{Value} \\
\hline
\zeta & \;\: 2.152 \\
\alpha_0 & \;\: 0.0002007 \\
\beta_0 & \;\: 0.003274 \\
\gamma_0 & \;\: 0.02187 \\
\varepsilon_0 & \;\: 0.0009816 \\
c & \;\: 1.492 \\
\theta_1 & \;\: 5.090 \\
\theta_2 & \;\: 2.718 \\
\kappa & \;\: 1.495 \\
\lambda_t(M_c) & \;\: 0.6057 \\
\tan \beta & 25.42 \\
\ea
$$
\end{table}

Our fit to the data is in Table \ref{t:ChiResults} and the corresponding Yukawa parameters for this
fit can be found in Table \ref{t:FitParams}.
A true $\chi^2$ function assumes gaussian errors for its observables, something which we have implicitly assumed but which is
not true for some of the observables used. Our $\chi^2$ is more of an indication of how good the fit is,
and the minimization of the function is a method by which we can find a ``best" set of parameters for the fit.
The fit value for our $\chi^2$ is around 5, with the majority of the contribution coming from the down-type quark sector.

The value for $m_u/m_c$ in the numerical fit ($0.0040$) is about 0.5$\sigma$ away from the value found in the
analytic section ($0.0037 \pm 0.0006$), and is consistent with our prediction of a larger $m_u$ and/or smaller
$m_c$ than measured.   In addition, the best fit value for $m_d m_s m_b$ in the $\chi^2$ analysis ($5.0 \times
10^5 \MeV^3$) is consistent with the value from the previous section ($(10.7 \pm 5.2) \times 10^5 \MeV^3$)
(within 1.1$\sigma$) and with the measured value ($(5.2 \pm 2.6) \times 10^5 \MeV$).  As for the free parameters,
in the analytic section we fit only one of these: $\l_t(M_c) = 0.59 \pm 0.12$. The numerical fit value found
here, $0.6057$, falls well within the range of the analytic fit value.  Thus, the results of the numerical
(Section \ref{sec:Numerical}) and analytic (Section \ref{sec:Analytic}) analyses are consistent.

\section{Justification in 6D \label{sec:6D} }

In this section we discuss a 6D version of our theory
which naturally justifies our setup given in this paper.
Our 5D analysis has assumed that the PS brane keeps its symmetry
even though $\chi^c$ and $\overline{\chi^c}$ get their VEVs near the cutoff scale
and break the PS symmetry entirely.
As mentioned in section \ref{sec:Setup}, there are generic large corrections
in the Kahler potentials of PS-brane fields $\psi$
\bea
K & = & (1 + C \f{\chi^{c\dagger}
\chi^c}{M_*^2} ) \psi^{\dagger} \psi. \nn
\eea
This problem can be solved in a geometric way by the addition of
a sixth dimension along which $\chi^c$ and $\psi$ are separated.
Even for a tiny sixth dimension, $1/R_2 \sim M_*/3$,
PS breaking effects are suppressed by $e^{-3}$ which is
enough suppression to keep our Yukawa relations.
All of the numerical analysis given
in the paper remains the same as long as the sixth dimension is not
too large compared to the cutoff scale.

There are two distinct ways of constructing 6D models.
First, a 6D N=1 (4D N=2) theory with $T^2/Z_2$ gives
four fixed points (3+1 dimensional spacetime).
All the fields in the bulk in the 5D theory are now in 6D
in the 6D theory, while all the fields living on branes
still live on branes.
As we have one 10 dimensional hypermultiplet
and two 16 dimensional hypermultiplets,
we need one additional 16 dimensional hypermultiplet
in order to cancel the 6D irreducible gauge anomaly of SO(10) \cite{Asaka:2002my}.

Second, a 6D N=1 theory with $T^2/(Z_2 \times Z'_2)$
also gives a 4D N=1 theory below the compactification scale.
There are four fixed lines along $x^5= 0, x^5 = \pi R_1/2, x^6= 0,$ and $x^6 = \pi R_2/2$
and four fixed points at the corners.
The gauge sector is extended to 6D,
but all of the other bulk fields can still be on 5D fixed lines.
This setup is anomaly free without introducing additional states.

Gauge coupling unification restricts the possible sizes of $R_1$ and $R_2$ in both cases. However, we can easily
recover the 5D theory used in this paper in the limit $1/R_2 \rightarrow M_*$.  Note, as long as $1/R_2$ is close
to, but somewhat less than, $M_*$  we simultaneously have gauge coupling unification (as described in the 5D
formulation) and PS symmetry relations for Yukawa couplings.

\begin{figure}[t]
\begin{center}

\begin{picture}(400,200)(0,0)

\CCirc(50,50){2}{Black}{Black}
\CCirc(350,50){2}{Black}{Black}
\CCirc(350,150){2}{Black}{Black}
\CCirc(50,150){2}{Black}{Black}

\Line(50,50)(350,50)
\Line(350,50)(350,150)
\Line(350,150)(50,150)
\Line(50,150)(50,50)

\Text(55,55)[bl]{SO(10)}
\Text(345,55)[br]{PS}
\Text(345,145)[tr]{SU(5)$^\prime \times $U(1)$_X^\prime$}
\Text(55,145)[tl]{SU(5)$\times$U(1)$_X$}

\SetColor{Blue}

\Line(50,25)(350,25)
\ArrowLine(349,25)(350,25)
\Text(200,15)[tc]{$x_5 \in \left[0 , \f{\pi R_1}{2} \right]$}

\Line(25,50)(25,150)
\ArrowLine(25,149)(25,150)
\rText(15,100)[rc][l]{$x_6 \in \left[0 , \f{\pi R_2}{2} \right]$}

\end{picture}

\caption{A representation of the 6D orbifold space.
The orbifold fixed points are at the corners of the space.
At each corner a different subgroup of SO(10) is preserved.
$x_5$ runs along the horizontal
direction from $0$ to $\f{\pi R_1}{2}$, while $x_6$ runs along the vertical direction
from $0$ to $\f{\pi R_2}{2}$.}
\label{f:6Dspace}
\end{center}
\end{figure}
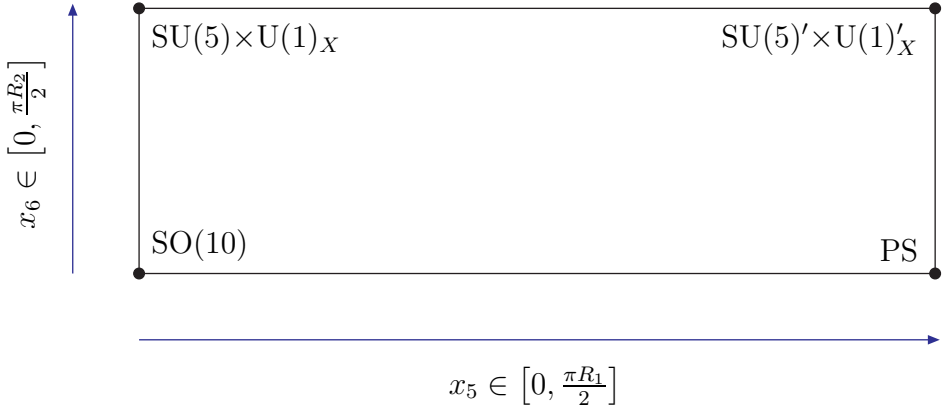

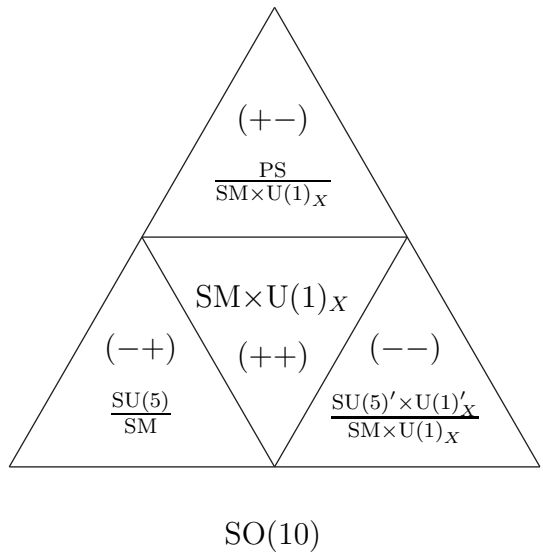
\begin{figure}[t]
\begin{center}

\begin{picture}(300,250)(0,0)

\Line(50,50)(250,50)
\Line(250,50)(150,223.31)
\Line(150,223.31)(50,50)

\Line(150,50)(200,136.66)
\Line(200,136.66)(100,136.66)
\Line(100,136.66)(150,50)

\Text(150,175.53)[bc]{$(+-)$}
\Text(150,165.53)[tc]{$\f{{\rm
PS}}{{\rm SM}\times{\rm U(1)}_X}$}

\Text(150,107.79)[bc]{SM$\times$U(1)$_X$}
\Text(150,97.79)[tc]{$(++)$}

\Text(100,88.87)[bc]{$(-+)$}
\Text(100,78.87)[tc]{$\f{{\rm
SU(5)}}{{\rm SM}}$}

\Text(200,88.87)[bc]{$(--)$}
\Text(200,78.87)[tc]{$\f{{\rm SU(5)}^\prime \times {\rm U(1)}^\prime_X}{ {\rm SM}\times{\rm U(1)}_X}$}

\Text(150,30)[tc]{SO(10)}

\end{picture}

\caption{A representation of the SO(10) group space. The space has been broken
into four subspaces by the orbifold symmetry, under which each subspace
has a different parity.}
\label{f:6Dgroupspace}
\end{center}
\end{figure}

We choose to focus on a 6D N=1 theory with $T^2/Z_2$. Such a theory has been studied by Asaka,
Buchmuller and Covi \cite{Asaka:2001eh} where the SO(10) gauge group in the bulk is
broken down to ${\rm SM} \times {\rm U(1)}_X$ by two different Wilson
lines. One breaks SO(10) down to the Pati-Salam gauge group along
$x_5$, and the other breaks SO(10) down to ${\rm SU(5)} \times {\rm U(1)}_X$
along $x_6$. Therefore, we end up with four fixed points; SO(10)
brane at $(x_5,x_6)=(0,0)$, Pati-Salam brane at $(\f{\pi R_1}{2},0)$, Georgi-Glashow
brane $({\rm SU(5)} \times {\rm U(1)}_X)$ at $(0,\f{\pi R_2}{2})$ and flipped
SU(5) brane $({\rm SU(5)}^{\prime} \times {\rm U(1)}_X^{\prime})$ at $(\f{\pi
R_1}{2},\f{\pi R_2}{2})$. We have provided a schematic of the extra-dimensional space
in Figure \ref{f:6Dspace}. The 5D setup we have considered so far can be
easily lifted up to this configuration. The procedure is the
following.
\begin{itemize}
\item 5D bulk states are extended to 6D bulk states. Extra
hypermultiplets are introduced to cancel the 6D anomaly.

\item Fields on the 5D SO(10) brane are located on the 6D SO(10) brane.

\item Fields on the 5D Pati-Salam brane are located on the 6D Pati-Salam brane except $\chi^c$ and
$\overline{\chi^c}$.  Note, in this framework we no longer need $\chi^c$ and $\overline{\chi^c}$, since the gauge
group is broken down to ${\rm SU(3)}_C \times {\rm SU(2)}_L \times {\rm U(1)}_Y \times {\rm U(1)}_X$ by the two
Wilson lines.  Nevertheless, any PS breaking effects on the SO(10) or PS branes, emanating from the PS breaking
fixed points, are suppressed. U(1)$_X$ is broken near the compactification scale by $16_{\rm kink}$ and
$\overline{\chi^c}_{\rm kink}$, the fields which generate the nonzero kink mass.
\end{itemize}

The 6D model is simple and economical. The additional states
introduced to cancel the 6D anomaly are hypermultiplets and can become
heavy by themselves. In addition they do not affect the differential running
of the three gauge couplings. Let us focus on gauge coupling
unification. The spectrum of massive Kaluza-Klein vector
multiplets are given by
\bea
M_{++} & =  \sqrt{ (\f{2n}{R_1})^2 + (\f{2m}{R_2})^2} & \;\; \left({\rm SM} \times {\rm U(1)}_X \right) \nn \\
M_{+-} & =  \sqrt{ (\f{2n}{R_1})^2 + (\f{2m+1}{R_2})^2} & \;\; \left(\f{{\rm PS}}{ {\rm SM}\times{\rm U(1)}_X} \right) \nn \\
M_{-+} & =  \sqrt{ (\f{2n+1}{R_1})^2 + (\f{2m}{R_2})^2} & \;\; \left(\f{{\rm SU(5)}}{{\rm SM}} \right) \nn \\
M_{--} & =  \sqrt{ (\f{2n+1}{R_1})^2 + (\f{2m+1}{R_2})^2} & \;\;
\left(\f{{\rm SU(5)}^\prime \times {\rm U(1)}^\prime_X}{ {\rm
SM}\times{\rm U(1)}_X} \right). \nn \eea The breakdown of SO(10)
into these subgroups is illustrated in Figure
\ref{f:6Dgroupspace}. If we send $R_2 \rightarrow 0$, we recover a
5D theory in which ${\rm SU(5)} \times {\rm U(1)}_X$ is broken to
the ${\rm SM} \times {\rm U(1)}_X$. The $R_2 \rightarrow 0$ limit
gives the same result as in Hall and Nomura \cite{Hall:2001xb}
(and also in Kim and Raby \cite{Kim:2002im}). Thus we fix $M_c
\sim 1/R_1$ at around $10^{14} \GeV$ with $M_* \sim 10^{17} \GeV$.
If $R_2 M_* \sim 2$ or 3, we get a tiny correction from extra $-+$
states and the result would be proportional to $\f{1}{3} ( b_{\rm
PS} - b_{\rm SM} ) \log (\f{R_2} {M_*}) $ which is neglegibly
small.\footnote{The notation used is defined in Kim and Raby
\cite{Kim:2002im}.} The elongated 6D rectangular configuration
gives a perfect setup for the construction of a realistic SO(10)
model.

The only remaining question is the applicability of bulk field
localization in 6D which was possible with a kink mass in 5D. Though the
quantitative results of 6D localization are different from the 5D case,
all of the qualitative aspects remain the same \cite{Lee:2003mc}.
More interestingly, 4D N=1 supersymmetry is preserved.

\section{Summary and Discussion \label{sec:Summary} }

Extra dimensions provide a nice framework for understanding how grand unified theories may be realized in nature.
In this paper we have constructed a 5D SO(10) model which accommodates the quark and charged lepton masses and
the CKM matrix. The model uses 11 parameters to fit the 13 independent observables of the quark and charged
lepton Yukawa sectors, allowing us to make two predictions: $\f{m_u}{m_c} (M_Z) = 0.0037 \pm 0.0006$ and $m_d m_s
m_b (M_Z) = (10.7 \pm 5.0) \times 10^5 \MeV^3$, both of which are roughly $1 \sigma$ larger than the experimental
values. The kink mass localizes the bulk Higgs fields according to their U(1)$_X$ quantum numbers, giving some
explanation for the hierarchy $m_b \ll m_t$ and $m_d > m_u$.  Our 5D SO(10) model can be considered to be an
effective theory coming from 6D SO(10) with one small and one large extra dimension. If the size of the sixth
dimension is very small (i.e. if the inverse of its characteristic length is near the cutoff scale) then all of
the calculations done here for the 5D model can be regarded as good approximations for the 6D case.

In a theory in which there are additional (heavy) vector-like states, commonly called Froggatt-Nielsen (FN)
fields, which have the same quantum numbers as the light states, the light and heavy states can mix. The extra
dimension naturally gives us a chance to unify Froggatt-Nielsen fields with ordinary matter fields if both come
from the same hypermultiplet in higher dimensions. In our model, however, we chose to place the 1st and 3rd
family fields on opposite branes to easily take advantage of the effects of the kink mass, and so our model does
not achieve the unification of the massless and FN states into bulk fields. Nevertheless, it may be possible to
place all of the matter fields in the bulk and still restrict the Yukawa terms of those fields to opposite branes
to gain the desired hierarchy from the kink mass. We leave this possibility for further research.

Further work is needed in six main areas in order to make our model complete. First, the neutrino sector should
be included to explain neutrino oscillation experiments. We have already fixed the neutrino Dirac masses in terms
of the Yukawa matrix between the left and right-handed neutrinos, but the heavy Majorana masses of the
right-handed neutrinos have not yet been determined. It would be very interesting to expand the model to include
Majorana masses and to investigate the resulting neutrino masses and mixings. Second, the weak scale
supersymmetry breaking mechanism should be specified. As there is as yet no such universally accepted mechanism,
we chose not to specify how SUSY is broken in our model. As a consequence, we could not calculate the electroweak
threshold corrections resulting from superparticle spectrum. Extra dimensions provide new interesting channels
for the understanding of supersymmetry breaking and its mediation and can give entirely different superparticle
spectra. We leave the weak scale supersymmetry breaking physics to future work. Third, our analysis is somewhat
incomplete since we have taken values for $\a_{\rm GUT}, M_G$ and $\ve_3$ from other sources. Hence our treatment
of gauge and Yukawa coupling RG running is not completely self consistent.   We have however accounted for this
shortcoming by including a theoretical uncertainty obtained by varying the gauge coupling parameters at $M_G$.
While a more complete unification treatment would be preferable, this would require knowledge of supersymmetry
breaking and the sparticle spectrum. Fourth, we have neglected the effects of running on the Yukawa matrices
between the compactification and cutoff scales. Our assumption is that such effects would contribute at the level
of a few percent. Further research is necessary to investigate these effects. Fifth, the UV completion of the
higher dimensional gauge theory could give us a better understanding of the model. String theory does not allow
arbitrary matter configurations and the constraints coming from it are usually stronger than those from field
theory. Therefore, it would be interesting to see if the 5D model considered here can be derived as an effective
field theory from a string theoretic starting point \cite{Kobayashi:2004ud,Kobayashi:2004ya}. Finally, we have
assumed that many fields in our theory obtain vacuum expectation values at certain scales and in particular
directions in group space. A complete theory would need to justify these VEVs by the minimization of the
appropriate potentials. Such work would be most easily done after finding a successful UV completion for the
theory, as many of these VEVs are of the order the cutoff energy scale.

\section*{Acknowledgments}

We thank the Institute for Advanced Study School of Natural Sciences for their hospitality,
and A. Falkowski for discussions on 5D orbifold propagators.
Partial support for this work came from DOE grant\# DOE/ER/01545-861.

\appendix

\section*{Appendix}

\section{D$_3$ Family Symmetry \label{sec:AppD_3}}

In this section, we define the D$_3$ group, give its character table, and give other information necessary to understand the couplings
and representations used in our model. Our presentation is an abbreviation of a more complete description given in \cite{Dermisek:1999vy}.

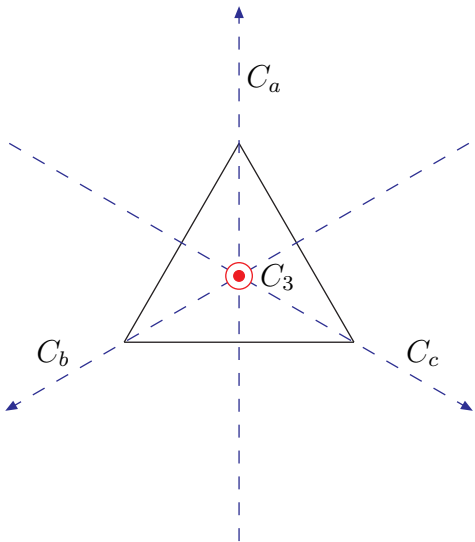
\begin{figure}[t]
\begin{center}

\begin{picture}(200,200)(0,0)

\SetColor{Black}
\Line(100,150)(56.70,75)
\Line(56.70,75)(143.30,75)
\Line(143.30,75)(100,150)

\SetColor{Blue}
\DashLine(100,0)(100,200){5}
\ArrowLine(100,199)(100,200)
\Text(110,175)[]{$C_a$}
\DashLine(186.60,150)(13.4,50){5}
\ArrowLine(14.266,50.5)(13.4,50)
\Text(30.05,71.16)[]{$C_b$}
\DashLine(13.4,150)(186.60,50){5}
\ArrowLine(185.734,50.5)(186.60,50)
\Text(169.95,71.16)[]{$C_c$}

\SetColor{Red}
\CCirc(100,100){5}{Red}{White}
\CCirc(100,100){2}{Red}{Red}
\Text(115,100)[]{$C_3$}

\end{picture}

\caption{Graphic representation of our choice of D$_3$ group elements. The solid (black) lines represent the triangle.
The 3 dashed (blue) lines represent the axes about which we take the $C_a$, $C_b$, and $C_c$ rotations.
The center (red) circle and dot represent the axis about which we take the $C_3$ rotation (it points out of the page).
Our rotation convention is right-handed.}
\label{f:D3pic}
\end{center}
\end{figure}

D$_3$ is the group of all rotations in three dimensions which leave an equilateral triangle invariant. The group contains six elements in
three classes: $E$; $C_3$, $C_3^2$; $C_a$, $C_b$, $C_c$. $E$ is the identity element. $C_3$ and $C_3^2$ signify $120^{\circ}$ and $240^\circ$
rotations about an axis through the center of and perpendicular to the triangle. $C_a$, $C_b$, and $C_c$ signify $180^\circ$ rotations about the three
different axes which run from the center to the three vertices of the triangle. We choose orientations such that $C_b = C_a C_3$ and $C_c = C_a C_3^2$.
Figure \ref{f:D3pic} contains a graphic representation of our conventions for the group elements.
The group has two inequivalent singlet representations:
${\bold 1}_{\rm A}$ and ${\bold 1}_{\rm B}$, and one doublet representation: ${\bold 2}_{\rm A}$.

The character table is
\bea
\ba{|c|c|c|c|}
\hline
{\rm D}_3 & E & C_3 & C_a \\
\hline
\; {\bold 1}_{\rm A} \; & \;\; 1 \;\; & \; 1 \; & \; 1 \; \\
\hline
\; {\bold 1}_{\rm B} \; & \; 1 \; & \; 1 \; & \; -1 \; \\
\hline
\; {\bold 2}_{\rm A} \; & \; 2 \; & \; -1 \; & \; 0 \; \\
\hline
\ea
\eea

We choose our representations as follows. When acting on a one dimensional representation, the elements are the characters
in the character table. When acting on the two dimensional representation, the elements are
\bea
E = \left(
\ba{cc}
1 & 0 \\
0 & 1
\ea
\right) \;\;\;\;
C_3 = \left(
\ba{cc}
\e & 0 \\
0 & \e^{-1}
\ea
\right) \;\;\;\;
C_a = \left(
\ba{cc}
0 & 1 \\
1 & 0
\ea
\right),
\eea
with $\e \equiv e^{2 \pi i/3}$. The remaining elements can be found by group multiplication.

Our Lagrangian of Section \ref{sec:Yukawa} contains terms with various combinations of D$_3$ fields. It is understood that the Lagrangian only contains
the D$_3$ singlet (${\bold 1}_{\rm A}$) part of these combinations. Here we list the combinations of fields we have used, and their singlet
projections.

Let $\t_i$ be ${\bold 2}_{\rm A}$ (doublet) fields under D$_3$, and let the internal degrees of freedom be represented by $x_i$ and $y_i$:
$\t_i \equiv
\left(
\ba{c}
x_i \\ y_i
\ea
\right)$.
Also, let $\phi$ be a ${\bold 1}_{\rm B}$ (``anti-symmetric" singlet) field, with internal degree of freedom $\a$: $\phi \equiv \a$.
Then the combinations we need are:
\bea
\left. \t_1 \times \t_2 \right|_{{\bold 1}_{\rm A}} & = & x_1 y_2 + y_1 x_2 \\
\left. \t_1 \times \t_2 \times \phi \right|_{{\bold 1}_{\rm A}} & = & (x_1 y_2 - y_1 x_2) \a \nn \\
\left. \t_1 \times \t_2 \times \t_3 \right|_{{\bold 1}_{\rm A}} & = & x_1 x_2 x_3 + y_1 y_2 y_3 \nn
\eea
Of course, multiplication by a ``true" singlet ${\bold 1}_{\rm A}$ has no effect concerning the D$_3$ structure of the Lagrangian terms.

\section{Massless States and Wavefunctions} \label{sec:AppWave}

This section will show how we determine the overlap between the original Lagrangian states and the massless
states. We choose to illustrate this through a specific example: the determination of the overlap between
the Lagrangian fields and the 3rd family left-handed states in equations (\ref{eq:3rdleft}) of section \ref{sec:Yukawa}.

The relevant superpotential terms are (from equation (\ref{eq:W_2})):
\bea
W_2 & \supset & \left( \overline{\psi}_{++} \right)_a (\partial_y - m_X) \left( \psi_{--} \right)_a \\
&& + \left[ 2 \sigma \phi_a \left( \overline{\psi}_{++} \right)_a \psi_3 \right] \delta (y)
+ \left[ 2 \eta \left( \overline{\psi}_{++} \right)_a \chi_a \right] \delta(y - \f{\pi R}{2}) \nn
\eea
The equations of motion are:
\bea
\f{\partial W}{\partial \left( \overline{\psi}_{++} \right)_a} & = & 0 \Longrightarrow \\
0 & = & (\partial_y - m_X) \left( \psi_{--} \right)_a + 2 \delta(y) \sigma \f{\phi_a}{M_*} \sqrt{M_*} \, \psi_3
+ 2 \delta(y- \f{\pi R}{2}) \eta \sqrt{M_*} \, \chi_a. \nn
\eea
We have added factors of $M_*$ where necessary to ensure that we have unitless couplings $\sigma$ and $\eta$.
This equation is satisfied for the massless projections of these fields.
Away from the branes, $\left( \psi_{--} \right)_a \propto e^{m_X y}$.
Because $\left( \psi_{--} \right)_a$ is odd under both orbifold parities and because $e^{m_X y}$ is
not an odd function about either $y=0$ or $y=\f{\pi R}{2}$, the profile of $\left( \psi_{--} \right)_a$
must have discontinuities at both branes. These discontinuities can cancel the brane terms in the equations of
motion. Define the overlap between $\psi_3$ and the massless field $\psi^0_3$ as $\psi_3 \supset \widetilde{n}_L \psi^0_3$. We can
then solve for the other two fields:
\bea
\left( \psi_{--} \right)_a & \supset &
-\varepsilon_{--}(y) e^{\f{X_L \zeta y}{\pi R}} \left( \f{\phi_a}{M_*} \right) \sigma \sqrt{M_*} \, \widetilde{n}_L \psi^0_3 \\
\chi_a & \supset & - e^{\f{X_L \zeta}{2}} \left( \f{\phi_a}{M_*} \right) \f{\sigma}{\eta} \widetilde{n}_L \psi^0_3 \nn
\eea
What remains is to determine the normalization $\widetilde{n}_L$. We do this by requiring that the effective 4D field
$\psi^0_3$ have a canonical Kahler potential term. Listing the fields which contribute to this term (we have suppressed the
gauge field factors here):
\bea
&& \int\limits^{\f{\pi R}{2}}_0 \mathrm{d}y \left\{ \psi_3^{\dag} \psi_3 \delta(y)
+ \sum_a \left( \psi_{--} \right)_a^\dag \left( \psi_{--} \right)_a
+ \sum_a \chi_a^\dag \chi_a \delta(y-\f{\pi R}{2}) \right\} \\
& \supset & \psi^{0 \, \dag}_3 \psi^0_3 \left| \widetilde{n}_L \right|^2
\left\{ 1 + \sum_a \left[
\sigma^2 M_* \left( \f{\phi_a}{M_*} \right)^2 \int\limits^{\f{\pi R}{2}}_0 \mathrm{d}y \ e^{\f{2 X_L \zeta y}{\pi R}}
+ \f{\sigma^2}{\eta^2} \left( \f{\phi_a}{M_*} \right)^2 e^{X_L \zeta}
\right] \right\} \nn
\eea
The integral is
\bea
\int\limits^{\f{\pi R}{2}}_0 \mathrm{d}y \ e^{\f{2 X_L \zeta y}{\pi R}} = \f{1}{M_* \rho^2 n_L^2}
\eea
which implies that for canonical normalization:
\bea
1 \equiv \left| \widetilde{n}_L \right|^2 \left\{ 1 + \f{\sigma^2}{\rho^2} \sum_a \left( \f{\phi_a}{M_*} \right)^2
\left[ \f{1}{n_L^2} + \f{\rho^2}{\eta^2} e^{X_L \zeta} \right] \right\}.
\eea
Choose $\widetilde{n}_L$ to be real:
\bea
\widetilde{n}_L & \equiv & \f{1}{\sqrt{1+r \left[ \f{1}{n_L^2} + \f{\rho^2}{\eta^2} e^{X_L \zeta} \right]}}
\eea

The other mixings and normalizations listed in Section \ref{sec:Yukawa} are calculated in the same way.

\section{A Lifted 4D Model \label{sec:App4DModel}}

We present here a 5D version of a 4D model by Dermisek and Raby in \cite{Dermisek:1999vy}
which itself was based on prior works \cite{Blazek:1999ue, Barbieri:1997tu, Barbieri:1996ww, Blazek:1996yv}.
Our purpose is to illustrate how such a 4D model could be placed into a 5D context and what
advantages can be gained from the use of the extra direction.

In using the extra dimension, we have separated the fields on the PS brane from the important
mass parameter $M_\chi = M_0 (1 + \a X)$ which gives much of the distinction between the particle types.
There are matter fields on the PS brane, and bulk matter fields which mediate between these brane fields
and the SO(10) $M_\chi$ VEV in such a way to give us the desired Yukawa matrix elements.

The setup for this model is largely the same as that presented in Section \ref{sec:Setup}, save
for the following points. This model does not have a kink mass in the bulk;
the small up quark mass comes instead from an approximate left-right symmetry in the model.
Here we use a $B-L$ VEV on the PS brane instead of $T_{3R}$.
Other differences lie in the placement of the matter fields and the extra VEV fields.
As with the model presented in the main body of the paper, for the current model under discussion we also have in mind
a setup which includes a small 6th dimension (as detailed in Section \ref{sec:6D}) in order to separate fields on the PS-brane
from the PS-breaking VEVs. This allows us to preserve the PS Yukawa relations on the PS-brane below the cutoff scale.

\begin{table}
\caption{Bulk fields}
\label{t:appbulkfields}
$$
\ba{lccc}
\mbox{Field} \;\; & PS \mbox{ Symm} \;\;\;\; & {\rm D}_3 \mbox{ Symm} \\
\hline
16 =
\left(
\ba{c}
\psi_{++} \\
\psi^c_{+-}
\ea
\right) &
\left(
\ba{c}
(4,2,1) \\
(\overline{4}, 1 , 2 )
\ea
\right) &
{\bf 2_A} \\
\overline{16} =
\left(
\ba{c}
\overline{\psi}_{--} \\
\overline{\psi^c}_{-+}
\ea
\right) &
\left(
\ba{c}
(\overline{4},2,1) \\
(4, 1, 2)
\ea
\right) &
{\bf 2_A} \\[5mm]
\hline
16^\prime =
\left(
\ba{c}
\psi^\prime_{+-} \\
\psi^{c \, \prime}_{++}
\ea
\right) &
\left(
\ba{c}
(4,2,1) \\
(\overline{4}, 1 , 2 )
\ea
\right) &
{\bf 2_A} \\
\overline{16^\prime} =
\left(
\ba{c}
\overline{\psi^\prime}_{-+} \\
\overline{\psi^{c \, \prime}}_{--}
\ea
\right) &
\left(
\ba{c}
(\overline{4},2,1) \\
(4, 1, 2)
\ea
\right) &
{\bf 2_A} \\[5mm]
\hline
\widetilde{16} =
\left(
\ba{c}
\widetilde{\psi}_{-+} \\
\widetilde{\psi^c}_{--}
\ea
\right) &
\left(
\ba{c}
(4,2,1) \\
(\overline{4}, 1 , 2 )
\ea
\right) &
{\bf 2_A} \\
\overline{\widetilde{16}} =
\left(
\ba{c}
\overline{\widetilde{\psi}}_{+-} \\
\overline{\widetilde{\psi^c}}_{++}
\ea
\right) &
\left(
\ba{c}
(\overline{4},2,1) \\
(4, 1, 2)
\ea
\right) &
{\bf 2_A} \\[5mm]
\hline
\widetilde{16^\prime} =
\left(
\ba{c}
\widetilde{\psi^\prime}_{--} \\
\widetilde{\psi^{c \, \prime}}_{-+}
\ea
\right) &
\left(
\ba{c}
(4,2,1) \\
(\overline{4}, 1 , 2 )
\ea
\right) &
{\bf 2_A} \\
\overline{\widetilde{16^\prime}} =
\left(
\ba{c}
\overline{\widetilde{\psi^\prime}}_{++} \\
\overline{\widetilde{\psi^{c \, \prime}}}_{+-}
\ea
\right) &
\left(
\ba{c}
(\overline{4},2,1) \\
(4, 1, 2)
\ea
\right) &
{\bf 2_A} \\[5mm]
\hline
10 =
\left(
\ba{c}
H_{++} \\
H^c_{+-}
\ea
\right) &
\left(
\ba{c}
(1,2,2) \\
(6,1,1)
\ea
\right) &
{\bf 1_A} \\
\overline{10} =
\left(
\ba{c}
\overline{H}_{--} \\
\overline{H^c}_{-+}
\ea
\right) &
\left(
\ba{c}
(1,2,2) \\
(6,1,1)
\ea
\right)
&
{\bf 1_A}
\ea
$$
\end{table}

\begin{table}
\caption{PS Brane fields}
\label{t:appPSfields}
$$
\ba{lcc}
\mbox{Field} \;\; & PS \mbox{ Symm} \;\;\;\; & {\rm D}_3 \mbox{ Symm} \\
\hline
\psi &
(4,2,1) &
{\bf 2_A} \\
\psi^c &
(\overline{4},1,2) &
{\bf 2_A} \\
\psi_3 &
(4,2,1) &
{\bf 1_A} \\
\psi^c_3 &
(\overline{4},1,2) &
{\bf 1_A} \\
\phi &
(1,1,1) &
{\bf 2_A} \\
\widetilde{\phi} &
(1,1,1) &
{\bf 2_A} \\
A &
(1,1,1) &
{\bf 1_B} \\
A_{15} &
(1,1,3) &
{\bf 1_A} \\
\Phi_L &
(1,1,1) &
{\bf 1_A} \\
\Phi_R &
(1,1,1) &
{\bf 1_A}
\ea
$$
\end{table}

The states are placed as follows. In the bulk, we have 8 matter hypermultiplets, forming 4 doublets under D$_3$ and
transforming as 16s under SO(10). Each of these hypermultiplet doublets has a different parity under the orbifold.
The Higgs, as before, is contained inside a $10$ hypermultiplet of SO(10). These fields are listed in Table \ref{t:appbulkfields}.
On the SO(10) brane there is only the mass parameter $M_\chi$, which is a singlet under D$_3$ and is a mix of
singlet and $X$ under SO(10):  $M_\chi = \l_\chi \left( \f{\chi^c \overline{\chi^c}}{M_*} \right) = M_0 (1 + \a X)$.
On the Pati-Salam brane, we have three sets of left- and
right-handed matter fields. Two of these sets form a doublet under D$_3$. In addition, there are several extra fields:
$\phi_a$, $\widetilde{\phi}_a$, $A$, $A_{15}$, $\Phi_L$, and $\Phi_R$. Of these extra fields, those with subscript $a$ are D$_3$ doublets. The rest
are ${\bf 1_A}$ save $A$ which is ${\bf 1_B}$ under D$_3$. All of these extra fields get nonzero VEVs except for $\widetilde{\phi}_1$.
All extra fields are SO(10) singlets, save $A_{15}$ which is an SU(4) adjoint and gets a VEV
$\left< A_{15} \right> = \left< A^0_{15} \right> (B-L)$. The PS brane fields are listed in Table \ref{t:appPSfields}.

We choose the following superpotential:
\bea
W_1 & = & \delta(y - \f{\pi R}{2}) \left\{ \lambda_1 \psi_3 H \psi^c_3
+ \lambda_2 \psi_a H \left( \psi^{c \prime}_{++} \right)_a \Phi_R
+ \lambda_2 \left( \psi_{++} \right)_a H \psi^c_a \Phi_L \right\} \\
W_2 & = & \overline{16}_a \partial_y 16_a
+ \overline{16^\prime}_a \partial_y 16^\prime_a
+ \overline{\widetilde{16}}_a \partial_y \widetilde{16}_a
+ \overline{\widetilde{16^\prime}}_a \partial_y \widetilde{16^\prime}_a
+ \overline{10} \partial_y 10 \nn \\[1mm]
&& + \delta(y) \left\{ M_\chi \overline{\widetilde{16}}_a 16^\prime_a + M_\chi \overline{\widetilde{16^\prime}}_a 16_a \right\} \nn \\
&& + \delta(y - \f{\pi R}{2}) \left\{
\left( \overline{\widetilde{\psi^\prime}}_{++} \right)_a
\left[ \lambda_3 A_{15} \phi_a \psi_3
+ \lambda_4 A_{15} \widetilde{\phi}_a \psi_a
+ \lambda_5 A \psi_a \right] \Phi_L \right. \nn \\
&& \hspace{21mm} \left.
+ \left( \overline{\widetilde{\psi^c}}_{++} \right)_a
\left[ \lambda_3 A_{15} \phi_a \psi^c_3
+ \lambda_4 A_{15} \widetilde{\phi}_a \psi^c_a
+ \lambda_5 A \psi^c_a \right] \Phi_R
\right\} \nn
\eea

There are 12 U(1) symmetries associated with our superpotential. We choose to parametrize
these by allowing the following 12 fields to be charged each under different U(1)s:
$\psi_3$, $\psi^c_3$, $H^c_{+-}$, $M_\chi$, $\left( \widetilde{\psi^\prime}_{--} \right)_a$,
$\left( \widetilde{\psi^{c \prime}}_{-+} \right)_a$, $\left( \widetilde{\psi}_{-+} \right)_a$,
$\Phi_L$, $A_{15}$. After specifying the U(1) symmetries of the above fields, the U(1) transformations
of all other fields are uniquely defined.
There are in addition 2 $Z_2$ symmetries. First is the $Z_2$ orbifold parity under $y \rightarrow -y$. The transformations
of the bulk fields under this symmetry have already been defined. In addition, let the rest of the independent fields:
$\psi_3$, $\psi^c_3$, $M_\chi$, $\Phi_L$, $A_{15}$ be even under this symmetry.
The second $Z_2$ involves a sign ambiguity in the transformation of $\Phi_R$. Under a given symmetry,
if $\Phi_L \rightarrow e^{i \alpha} \Phi_L$, then the superpotential terms imply that
$\Phi_R \rightarrow e^{i n \pi} e^{i \alpha} \Phi_R$ with $n=0,1$. We choose to require that $n=1$ here in order
to forbid terms created by the replacement of $\Phi_L$ by $\Phi_R$ or vice-versa. We also assume that
$\left< \Phi_L \right> / M_* \ll 1$ so that replacements like $\Phi_L \rightarrow \Phi_R^2$ are negligible.\footnote{Such terms
given by the substitution $\Phi_L \rightarrow \Phi_R^2$ or $\Phi_R \rightarrow \Phi_L^2$ would lead to a Yukawa
matrix structure different from the one desired and so should be forbidden by some symmetry.}
Let the 12 independent fields be uncharged under this symmetry. We require all of these symmetries just listed
to be symmetries of the theory so as to forbid unwanted extra terms in the superpotential.

The superpotential has a left-right symmetry, under which
$\psi_3 \leftrightarrow \psi^c_3$, $\psi \leftrightarrow \psi^c$,
$ \left( \ba{c} 16 \\ \overline{16} \ea  \right) \leftrightarrow \left( \ba{c} 16^\prime \\ \overline{16^\prime} \ea  \right) $,
$ \left( \ba{c} \widetilde{16} \\ \overline{\widetilde{16}} \ea  \right)
 \leftrightarrow \left( \ba{c} \widetilde{16^\prime} \\ \overline{\widetilde{16^\prime}} \ea  \right)$,
 $\Phi_L \leftrightarrow \Phi_R$.
This symmetry, which commutes with the family D$_3$ symmetry,
is broken spontaneously by the VEVs of $\Phi_L$ and $\Phi_R$, which we require to be slightly different.
This difference is encapsulated by a small parameter $\eta$: $\left< \Phi_R \right> = \left< \Phi_L \right> (1 + \eta)$.

We now turn to the equations of motion for the left-handed states:
\bea
\f{\partial W}{\partial_y \left( \overline{\psi}_{--} \right)_a} = 0 & \Longrightarrow & \partial_y \left( \psi_{++} \right)_a = 0 \\
\f{\partial W}{\partial_y \left( \overline{\psi^\prime}_{-+} \right)_a} = 0 & \Longrightarrow & \partial_y \left( \psi^\prime_{+-} \right)_a = 0 \nn \\
\f{\partial W}{\partial_y \left( \overline{\widetilde{\psi}}_{+-} \right)_a} = 0
& \Longrightarrow & \partial_y \left( \widetilde{\psi}_{-+} \right)_a + \delta(y) M_\chi \left( \psi^\prime_{+-} \right)_a = 0 \nn \\
\f{\partial W}{\partial_y \left( \overline{\widetilde{\psi^\prime}}_{++} \right)_1} = 0
& \Longrightarrow & \partial_y \left( \widetilde{\psi^\prime}_{--} \right)_2 + \delta(y) M_\chi \left( \psi_{++} \right)_2 \nn \\ \nn
&& + \delta(y - \f{\pi R}{2}) \left[ \lambda_3 A_{15} \phi_2 \psi_3 + \lambda_4 A_{15} \widetilde{\phi}_1 \psi_1 + \lambda_5 A \psi_2 \right] \Phi_L = 0 \nn \\
\f{\partial W}{\partial_y \left( \overline{\widetilde{\psi^\prime}}_{++} \right)_2} = 0
& \Longrightarrow & \partial_y \left( \widetilde{\psi^\prime}_{--} \right)_1 + \delta(y) M_\chi \left( \psi_{++} \right)_1 \nn \\ \nn
&& + \delta(y - \f{\pi R}{2}) \left[ \lambda_3 A_{15} \phi_1 \psi_3 + \lambda_4 A_{15} \widetilde{\phi}_2 \psi_2 - \lambda_5 A \psi_1 \right] \Phi_L = 0 \nn
\eea
Solving these equations leads to knowledge of the overlap between the massless fields and the original fields in the
Lagrangian. The following relationships only have the massless components on the right hand sides of the equations.
(We have replaced the brane fields by their VEVs)
\bea
\left( \psi^\prime_{+-} \right)_a & \supset & 0 \\
\left( \widetilde{\psi}_{-+} \right)_a & \supset & 0 \nn \\
\left( \widetilde{\psi^\prime}_{--} \right)_1 & \supset & \f{1}{2} \widetilde{\varepsilon}(y)
\left[ \lambda_3 \left< A_{15} \right> \left< \phi_1 \right> \psi_3 + \lambda_4 \left< A_{15} \right> \left< \widetilde{\phi}_2 \right> \psi_2
- \lambda_5 \left< A \right> \psi_1 \right] \left< \Phi_L \right> \nn \\
\left( \widetilde{\psi^\prime}_{--} \right)_2 & \supset & \f{1}{2} \widetilde{\varepsilon}(y)
\left[ \lambda_3 \left< A_{15} \right>  \left< \phi_2 \right> \psi_3 + \lambda_5 \left< A \right> \psi_2 \right] \left< \Phi_L \right> \nn \\
\left( \psi_{++} \right)_1 & \supset & -\f{1}{M_\chi}
\left[ \lambda_3 \left< A_{15} \right> \left< \phi_1 \right> \psi_3 + \lambda_4 \left< A_{15} \right> \left< \widetilde{\phi}_2 \right> \psi_2
- \lambda_5 \left< A \right> \psi_1 \right] \left< \Phi_L \right> \nn \\
\left( \psi_{++} \right)_2 & \supset & -\f{1}{M_\chi}
\left[ \lambda_3 \left< A_{15} \right> \left< \phi_2 \right> \psi_3 + \lambda_5 \left< A \right> \psi_2 \right] \left< \Phi_L \right> \nn
\eea

The equations for the right-handed states can be obtained from the left-right symmetry present in the model.
We list the most important of these equations:
\bea
\left( \psi^{c \prime}_{++} \right)_1 & \supset & -\f{1}{M_\chi}
\left[ \lambda_3 \left< A_{15} \right> \left< \phi_1 \right> \psi^c_3 + \lambda_4 \left< A_{15} \right> \left< \widetilde{\phi}_2 \right> \psi^c_2
- \lambda_5 \left< A \right> \psi^c_1 \right] \left< \Phi_R \right> \\
\left( \psi^{c \prime}_{++} \right)_2 & \supset & -\f{1}{M_\chi}
\left[ \lambda_3 \left< A_{15} \right> \left< \phi_2 \right> \psi^c_3 + \lambda_5 \left< A \right> \psi^c_2 \right] \left< \Phi_R \right> \nn
\eea

The three $\psi_i$ fields span the space of the left-handed massless states. Because the other fields which we've integrated out have
massless components, the kinetic energy and gauge interaction terms for the $\psi_i$ fields are no longer orthonormal.
The same is true for the $\psi^c_i$ states by the left-right symmetry.
We will discuss the effects of rotating and rescaling these fields to an orthonormal basis later.

We replace the $\left( \psi_{++} \right)_a$ and $\left( \psi^{c \prime}_{++} \right)_a$ fields in $W_1$ by their corresponding
massless parts in order to get the low energy Yukawa matrices. The result:\footnote{We have chosen to use the notation
found in \cite{Dermisek:1999vy} to ease comparison between prior works and our own.}
\bea
Y_u & = &
\left(
\ba{ccc}
0 & \varepsilon^\prime \rho & r \varepsilon \kappa T_{u^c} \\
-\varepsilon^\prime \rho & \varepsilon \rho & r \varepsilon T_{u^c} \\
r \varepsilon \kappa T_Q & r \varepsilon T_Q & 1
\ea
\right) \lambda \\
Y_d & = &
\left(
\ba{ccc}
0 & \varepsilon^\prime & r \varepsilon \sigma \kappa T_{d^c} \\
-\varepsilon^\prime & \varepsilon & r \varepsilon \sigma T_{d^c} \\
r \varepsilon \kappa T_Q & r \varepsilon T_Q & 1
\ea
\right) \lambda \nn \\
Y_e & = &
\left(
\ba{ccc}
0 & -\varepsilon^\prime & r \varepsilon \kappa T_{e^c} \\
\varepsilon^\prime & 3 \varepsilon & r \varepsilon T_{e^c} \\
r \varepsilon \sigma \kappa T_L & r \varepsilon \sigma T_L & 1
\ea
\right) \lambda \nn
\eea

Definitions follow for these variables, where we have used $M_\chi = M_0 (1 + \alpha X)$,
$\left< \Phi_R \right> = \left< \Phi_L \right> (1 + \eta)$, and have added factors of the cutoff scale in order
to make the couplings $\lambda_i$ all unitless. We have also assumed $\eta \ll 1$ and $\alpha \sim {\cal O} (1)$.
$T_f$ represents the $(B-L)$ quantum number for the field $f$.
\bea
\varepsilon & \equiv & \f{\lambda_2 \lambda_4 \left< A^0_{15} \right> \left< \widetilde{\phi}_2 \right> \left< \Phi_L \right>^2 }{3 \lambda_1 M_*^3 M_0}
\f{4 \alpha}{(1+\alpha)(1-3 \alpha)} \\
\varepsilon^\prime & \equiv & - \f{\lambda_2 \lambda_5 \left< A \right> \left< \Phi_L \right>^2}{\lambda_1 M_*^2 M_0}
\f{4 \alpha}{(1+\alpha)(1-3 \alpha)} \nn \\
\rho & \equiv & \f{2 \eta (1-3 \alpha)}{4 \alpha} \nn \\
r & \equiv & - \f{\lambda_3 \left< \phi_1 \right>}{\lambda_4 \left< \widetilde{\phi}_2 \right>} \f{3 (1-3 \alpha)}{4 \alpha} \nn \\
\kappa & \equiv & \f{\left< \phi_2 \right>}{\left< \phi_1 \right>} \nn \\
\sigma & \equiv & \f{1+\alpha}{1-3 \alpha} \nn \\
\lambda & \equiv & \lambda_1 \sqrt{\f{2 M_c}{\pi M_*} } \nn
\eea
These Yukawa matrices are the same as those found in \cite{Dermisek:1999vy}, except for the (1,3) and (3,1) elements. It has been shown in
\cite{Kim:2004ki} that (1,3) elements are needed in models of this kind in order to fit $\sin 2 \b$.

Here we list the lowest-order diagrams which give the Yukawa matrices. Each diagram is followed by the element(s)
to which it contributes.

\newpage

\begin{center}
\SetPFont{Helvetica}{15}
\begin{picture}(300,520)(0,0)

\SetColor{Blue} \ArrowLine(75,500)(150,500) \Text(112.5,485)[cc]{$\psi_3$}

\SetColor{Red}  \ArrowLine(150,520)(150,500) \Text(140,520)[cc]{$H$}

\SetColor{Blue} \ArrowLine(225,500)(150,500) \Text(187.5,485)[cc]{$\psi_3^c$}

\SetColor{Black} \Text(340,500)[cc]{(3,3)}

\SetColor{Blue} \ArrowLine(0,420)(75,420) \Text(37.5,405)[cc]{$\psi_a$}

\SetColor{Red}  \ArrowLine(60,440)(75,420) \Text(50,440)[cc]{$H$}

\SetColor{Red}  \ArrowLine(90,440)(75,420) \Text(110,440)[rc]{$\Phi_R$}

\SetColor{Black} \ArrowLine(150,420)(75,420) \Text(112.5,405)[cc]{$\left( \psi^{c \prime}_{++} \right)_a$}

\SetColor{Black} \Text(150,422)[cc]{X} \Text(150,435)[cc]{$M_\chi$}

\SetColor{Black} \ArrowLine(150,420)(225,420) \Text(187.5,405)[cc]{$\left( \overline{\widetilde{\psi^c}}_{++} \right)_a$}

\SetColor{Red} \ArrowLine(210,440)(225,420) \Text(190,440)[lc]{$A_{15}$}

\SetColor{Red} \ArrowLine(225,440)(225,420) \Text(225,450)[cc]{$\phi_a$}

\SetColor{Red} \ArrowLine(240,440)(225,420) \Text(260,440)[rc]{$\Phi_R$}

\SetColor{Blue} \ArrowLine(300,420)(225,420) \Text(262.5,405)[cc]{$\psi^c_3$}

\SetColor{Black} \Text(340,420)[cc]{(1,3) (2,3)}

\SetColor{Blue} \ArrowLine(0,340)(75,340) \Text(37.5,325)[cc]{$\psi_3$}

\SetColor{Red}  \ArrowLine(60,360)(75,340) \Text(40,360)[lc]{$A_{15}$}

\SetColor{Red} \ArrowLine(75,360)(75,340) \Text(75,370)[cc]{$\phi_a$}

\SetColor{Red}  \ArrowLine(90,360)(75,340) \Text(110,360)[rc]{$\Phi_L$}

\SetColor{Black} \ArrowLine(150,340)(75,340) \Text(112.5,325)[cc]{$\left( \overline{\widetilde{\psi^\prime}}_{++} \right)_a$}

\SetColor{Black} \Text(150,342)[cc]{X} \Text(150,355)[cc]{$M_\chi$}

\SetColor{Black} \ArrowLine(150,340)(225,340) \Text(187.5,325)[cc]{$\left( \psi_{++} \right)_a$}

\SetColor{Red} \ArrowLine(210,360)(225,340) \Text(200,360)[cc]{$H$}

\SetColor{Red} \ArrowLine(240,360)(225,340) \Text(260,360)[rc]{$\Phi_L$}

\SetColor{Blue} \ArrowLine(300,340)(225,340) \Text(262.5,325)[cc]{$\psi^c_a$}

\SetColor{Black} \Text(340,340)[cc]{(3,1) (3,2)}

\SetColor{Blue} \ArrowLine(0,260)(75,260) \Text(37.5,245)[cc]{$\psi_a$}

\SetColor{Red}  \ArrowLine(60,280)(75,260) \Text(50,280)[cc]{$H$}

\SetColor{Red}  \ArrowLine(90,280)(75,260) \Text(110,280)[rc]{$\Phi_R$}

\SetColor{Black} \ArrowLine(150,260)(75,260) \Text(112.5,245)[cc]{$\left( \psi^{c \prime}_{++} \right)_a$}

\SetColor{Black} \Text(150,262)[cc]{X} \Text(150,275)[cc]{$M_\chi$}

\SetColor{Black} \ArrowLine(150,260)(225,260) \Text(187.5,245)[cc]{$\left( \overline{\widetilde{\psi^c}}_{++} \right)_a$}

\SetColor{Red} \ArrowLine(210,280)(225,260) \Text(190,280)[lc]{$A_{15}$}

\SetColor{Red} \ArrowLine(225,280)(225,260) \Text(225,290)[cc]{$\widetilde{\phi_a}$}

\SetColor{Red} \ArrowLine(240,280)(225,260) \Text(260,280)[rc]{$\Phi_R$}

\SetColor{Blue} \ArrowLine(300,260)(225,260) \Text(262.5,245)[cc]{$\psi^c_a$}

\SetColor{Black} \Text(340,260)[cc]{(2,2)}

\SetColor{Blue} \ArrowLine(0,180)(75,180) \Text(37.5,165)[cc]{$\psi_a$}

\SetColor{Red}  \ArrowLine(60,200)(75,180) \Text(40,200)[lc]{$A_{15}$}

\SetColor{Red} \ArrowLine(75,200)(75,180) \Text(75,210)[cc]{$\widetilde{\phi_a}$}

\SetColor{Red}  \ArrowLine(90,200)(75,180) \Text(110,200)[rc]{$\Phi_L$}

\SetColor{Black} \ArrowLine(150,180)(75,180) \Text(112.5,165)[cc]{$\left( \overline{\widetilde{\psi^\prime}}_{++} \right)_a$}

\SetColor{Black} \Text(150,182)[cc]{X} \Text(150,195)[cc]{$M_\chi$}

\SetColor{Black} \ArrowLine(150,180)(225,180) \Text(187.5,165)[cc]{$\left( \psi_{++} \right)_a$}

\SetColor{Red} \ArrowLine(210,200)(225,180) \Text(200,200)[cc]{$H$}

\SetColor{Red} \ArrowLine(240,200)(225,180) \Text(260,200)[rc]{$\Phi_L$}

\SetColor{Blue} \ArrowLine(300,180)(225,180) \Text(262.5,165)[cc]{$\psi^c_a$}

\SetColor{Black} \Text(340,180)[cc]{(2,2)}

\SetColor{Blue} \ArrowLine(0,100)(75,100) \Text(37.5,85)[cc]{$\psi_a$}

\SetColor{Red}  \ArrowLine(60,120)(75,100) \Text(50,120)[cc]{$H$}

\SetColor{Red}  \ArrowLine(90,120)(75,100) \Text(110,120)[rc]{$\Phi_R$}

\SetColor{Black} \ArrowLine(150,100)(75,100) \Text(112.5,85)[cc]{$\left( \psi^{c \prime}_{++} \right)_a$}

\SetColor{Black} \Text(150,102)[cc]{X} \Text(150,115)[cc]{$M_\chi$}

\SetColor{Black} \ArrowLine(150,100)(225,100) \Text(187.5,85)[cc]{$\left( \overline{\widetilde{\psi^c}}_{++} \right)_a$}

\SetColor{Red} \ArrowLine(210,120)(225,100) \Text(200,120)[cc]{$A$}

\SetColor{Red} \ArrowLine(240,120)(225,100) \Text(260,120)[rc]{$\Phi_R$}

\SetColor{Blue} \ArrowLine(300,100)(225,100) \Text(262.5,85)[cc]{$\psi^c_a$}

\SetColor{Black} \Text(340,100)[cc]{(1,2) (2,1)}

\SetColor{Blue} \ArrowLine(0,20)(75,20) \Text(37.5,5)[cc]{$\psi_a$}

\SetColor{Red}  \ArrowLine(60,40)(75,20) \Text(50,40)[cc]{$A$}

\SetColor{Red}  \ArrowLine(90,40)(75,20) \Text(110,40)[rc]{$\Phi_L$}

\SetColor{Black} \ArrowLine(150,20)(75,20) \Text(112.5,5)[cc]{$\left( \overline{\widetilde{\psi^\prime}}_{++} \right)_a$}

\SetColor{Black} \Text(150,22)[cc]{X} \Text(150,35)[cc]{$M_\chi$}

\SetColor{Black} \ArrowLine(150,20)(225,20) \Text(187.5,5)[cc]{$\left( \psi_{++} \right)_a$}

\SetColor{Red} \ArrowLine(210,40)(225,20) \Text(200,40)[cc]{$H$}

\SetColor{Red} \ArrowLine(240,40)(225,20) \Text(260,40)[rc]{$\Phi_L$}

\SetColor{Blue} \ArrowLine(300,20)(225,20) \Text(262.5,5)[cc]{$\psi^c_a$}

\SetColor{Black} \Text(340,20)[cc]{(1,2) (2,1)}

\end{picture}
\end{center}

We performed a $\chi^2$ analysis, the same as that used in \cite{Blazek:1999ue}, save that here we have nonzero (1,3) and (3,1) terms and
a new parameter $\kappa$. The fit parameters follow.

First the GUT parameters:
\bea
\f{1}{\a_{\rm GUT}} & = & 25.12  \\
M_G & = & 2.54 \times 10^{16} \GeV \nn \\
\varepsilon_3 & = & -3.61 \% \nn
\eea
Next the (modulus) Yukawa sector:
\bea
\lambda & = & 0.70 \\
r & = & 24.8 \nn \\
\sigma & = & 1.19 \nn \\
\varepsilon & = & 0.0090 \nn \\
\rho & = & 0.061 \nn \\
\varepsilon^\prime & = & 0.0034 \nn \\
\kappa & = & 0.15 \nn
\eea
Next the Yukawa phase information ($\phi_x \equiv \arg (x) $) in radians:
\bea
\phi_\sigma & = & \;\;\: 0.51 \\
\phi_\rho & = & -1.87 \nn \\
\phi_\kappa & = & \;\;\: 0.89 \nn
\eea
The other Yukawa parameters are assumed to be real.
Finally some SUSY breaking scales and $\tan \b$:
\bea
\mu & = & 234 \GeV \\
M_{1/2} & = & 606 \GeV \nn \\
m_{16} & = & 4160 \GeV \nn \\
A_0 & = & -7736 \GeV \nn \\
\left( \f{m_{10}}{m_{16}} \right)^2 & = &  1.80 \nn \\
\left( \f{m_D}{m_{16}} \right)^2 & = & 0.128 \nn \\
\tan \beta & = & 52.2 \nn
\eea

\begin{table}
\caption{Best fit. Mass dimensions are in $\GeV$.}
\label{t:appfit}
$$
\ba{llll}
\mbox{Observable} \hspace{10mm} & \mbox{Target Value} \hspace{15mm} & \mbox{Fit Value} \hspace{5mm} & \chi^2 \mbox{ Contribution} \\
\hline
\f{1}{\a_{\rm EM}} &  137.04 \pm 0.14 & 137.0 & 0.08 \\
G_{\mu} \times 10^5 & 1.1664 \pm 0.0012 & 1.166 & 0.11 \\
\a_s & 0.1172 \pm 0.0020 & 0.1167 & 0.06 \\
M_{\rm Top} & 178.0 \pm 4.3 & 176.5 & 0.12 \\
m_b & 4.220 \pm 0.090 & 4.243 & 0.07 \\
M_b - M_c & 3.40 \pm 0.20 & 3.35 & 0.06 \\
m_s & 0.089 \pm 0.011 & 0.104 & 1.86 \\
\f{1}{Q^2} \times 10^3 & 2.03 \pm 0.20 & 2.00 & 0.02 \\
\f{m_d}{m_s} & 0.050 \pm 0.015 & 0.074 & 2.54 \\
M_\tau & 1.7770 \pm 0.0018 & 1.777 & 0.00 \\
M_\mu & 0.10566 \pm 0.00011 & 0.1057 & 0.13 \\
M_e \times 10^3 & 0.51100 \pm 0.00051 & 0.5110 & 0.00 \\
V_{us} & 0.2230 \pm 0.0040 & 0.2216 & 0.12 \\
V_{cb} & 0.0402 \pm 0.0019 & 0.0390 & 0.40 \\
\f{V_{ub}}{V_{cb}} & 0.0860 \pm 0.0080 & 0.0863 & 0.00 \\
\varepsilon_K \times 10^3 & 2.28 \pm 0.23 & 2.35 & 0.10 \\
M_Z & 91.188 \pm 0.091 & 91.19 & 0.00 \\
M_W & 80.419 \pm 0.080 & 80.41 & 0.01 \\
m_c & 1.30 \pm 0.15 & 1.15 & 1.00 \\
(b \rightarrow s \gamma) \times 10^3 & 0.334 \pm 0.038 & 0.335 & 0.00 \\
\sin 2 \beta & 0.727 \pm 0.036 & 0.700 & 0.57 \\
V_{td} \times 10^3 & 8.20 \pm 0.82 & 8.35 & 0.03 \\
\hline
& & \mbox{Total:} & 7.30
\ea
$$
\end{table}

The fit itself can be found in Table \ref{t:appfit}.
As explained in Section \ref{sec:Numerical}, we are using our $\chi^2$ function as a vehicle to find the best possible fit
rather than as a true statistical $\chi^2$ function. Our best $\chi^2$ is 7.30,
indicating that we are not fitting some of the data.
Specifically, as shown in Table \ref{t:appfit}, the observables $m_s$, $\f{m_d}{m_s}$, and $m_c$ have $\chi^2$ contributions
of 1 or greater and make up the majority of the $\chi^2$ fit value. The fit values of the down and strange quark masses are on the
large side, while the charm mass fit value is smaller than the data. We present this fit as a first attempt at this
kind of model. More work is needed to alter the model to obtain a better fit.

In our analysis we have used a basis in which the massless matter fields are not orthonormal. Rotation and rescaling
to a canonical orthonormal basis would in general introduce changes to the Yukawa matrices. We have neglected effects
from this change of basis, and our justification for this is the following. Were a fit to be done with these
effects included, the input Yukawa parameters would compensate by changing their values. We assume that the
input parameters could compensate to the extent that we would obtain essentially the same fit in this case as in the case
we have presented here in the appendix without these extra effects. We leave it to further research to explore whether
this assumption is a good one.

\end{document}